\documentclass[useAMS,usenatbib]{mn2e}
\usepackage{amssymb}
\usepackage{latexsym,graphicx}

\def\mnrasreferences{\longrefs=1 \bibliographystyle{mn2e}}

\newcount\longrefs
\def\aap{\ifnum\longrefs=1 {Astron.\ Astrophys.}\else 
                           {A\hbox{\rm \&}A}\fi}
\def\aapr{\ifnum\longrefs=1 {Astron.\ Astrophys.\ Rev.}\else 
                            {A\hbox{\rm \&}AR}\fi}
\def\aaps{\ifnum\longrefs=1 {Astron.\ Astrophys.\ Suppl.}\else 
                            {A\hbox{\rm \&}AS}\fi}
\def\aj{\ifnum\longrefs=1 {Astron.\ J.}\else 
                          {AJ}\fi} 
\def\ao{\ifnum\longrefs=1 {Applied Optics}\else 
                           {Appl.\ Opt.}\fi} 
\def\aspcs{\ifnum\longrefs=1 {Astron.\ Soc.\ Pacific Conf. Series}\else 
                           {ASP Conf.\ Ser.}\fi} 
\def\apj{\ifnum\longrefs=1 {Astrophys.\ J.}\else 
                           {ApJ}\fi} 
\def\apjl{\ifnum\longrefs=1 {Astrophys.\ J. Lett.}\else 
                            {ApJ}\fi} 
\def\aplett{\ifnum\longrefs=1 {Astrophys.\ J. Lett.}\else 
                            {ApJ}\fi} 
\def\apjs{\ifnum\longrefs=1 {Astrophys.\ J. Suppl.}\else 
                            {ApJS}\fi}
\def\apss{\ifnum\longrefs=1 {Astrophys.\ and Space Science}\else 
                            {Ap\hbox{\rm \&}SS}\fi}
\def\araa{\ifnum\longrefs=1 {Ann.\ Rev.\ Astron.\ Astrophys.}\else 
                            {ARA\hbox{\rm \&}A}\fi}
\def\azh{\ifnum\longrefs=1 {Astronomicheskii Zhurnal}\else 
                            {Astron.\ Zhur.}\fi}
\def\baas{\ifnum\longrefs=1 {Bull.\ Am.\ Astron.\ Soc.}\else 
                            {BAAS}\fi}
\def\bain{\ifnum\longrefs=1 {Bull.\ Astronom.\ Institutes Netherlands}\else
                            {Bull.\ Astr.\ Inst.\ Neth.}\fi}
\def\gca{\ifnum\longrefs=1 {Geochim.\ Cosmochim.\ Acta}\else 
                           {Geochim.\ Cosmochim.\ Acta}\fi}
\def\grl{\ifnum\longrefs=1 {Geophys.\ Res.\ Lett.}\else 
                           {Geoph.\ Res.\ Lett.}\fi}
\def\iaucirc{\ifnum\longrefs=1 {IAU Circulars}\else 
                          {IAU Circ.}\fi}
\def\ip{\ifnum\longrefs=1 {in press}\else 
                          {in press}\fi}
\def\jgr{\ifnum\longrefs=1 {J.\ Geophys.\ Res.}\else 
                           {J.\ Geophys.\ Res.}\fi}  
\def\jrasc{\ifnum\longrefs=1 {J.\ Royal Astron.\ Soc.\ Canada}\else 
                           {JRAS Can.}\fi}  
\def\mnras{\ifnum\longrefs=1 {Mon.\ Not.\ Roy.\ Astron.\ Soc.}\else 
                             {MNRAS}\fi} 
\def\nat{\ifnum\longrefs=1 {Nature}\else 
                           {Nat}\fi}
\def\pasj{\ifnum\longrefs=1 {Pub.\ Astron.\ Soc.\ Japan}\else 
                            {PASJ}\fi} 
\def\pasp{\ifnum\longrefs=1 {Pub.\ Astron.\ Soc.\ Pacific}\else 
                            {PASP}\fi} 
\def\physscr{\ifnum\longrefs=1 {Physica Scripta}\else 
                            {Phys.\ Scrip.}\fi} 
\def\planss{\ifnum\longrefs=1 {Planetary \& Space Science}\else 
                            {Plan. \& Space Sci.}\fi} 
\def\procspie{\ifnum\longrefs=1 {Proc.\ SPIE}\else 
                            {Proc.\ SPIE}\fi} 
\def\qjras{\ifnum\longrefs=1 {Quarterly J.\ Royal Astron.\ Soc.}\else 
                            {QJRAS}\fi} 
\def\sa{\ifnum\longrefs=1 {Soviet Astron..}\else 
                               {Sov.\ Astron.}\fi}
\def\skytel{\ifnum\longrefs=1 {Sky \& Telescope}\else 
                            {Sky \& Tel.}\fi} 
\def\solphys{\ifnum\longrefs=1 {Solar Phys.}\else 
                               {Solar Phys.}\fi}
\def\ssr{\ifnum\longrefs=1 {Space Science Rev.}\else 
                               {Space\ Sci.\ Rev.}\fi}

\def\dutch{\def\refname{Referenties}\def\abstractname{Samenvatting}%
  \def\bibname{Bibliografie}\def\chaptername{Hoofdstuk}%
  \def\appendixname{Bijlage}\def\contentsname{Inhoudsopgave}%
  \def\listfigurename{Lijst van figuren}\def\listtablename{Lijst van tabellen}%
  \def\indexname{Index}\def\figurename{Figuur}\def\tablename{Tabel}%
  \def\partname{Deel}\def\enclname{Bijlage(n)}\def\ccname{Ter attentie van}%
  \def\headtoname{Aan}\def\headpagename{Pagina}%
  \def\today{\number\day\space\ifcase\month\or januari\or februari\or maart\or%
     april\or mei\or juni\or juli\or augustus\or september\or oktober\or%
     november\or december\fi \space\number\year}%
  \typeout{
              >>>>> use hlatex209 for Dutch hyphenation <<<<< 
         }}
\newcounter{onefig} \newcounter{fignumber}
\newcount\nocaptions \newcount\nofigures \newcount\figwidth
\newcount\viewgraphs
  \def\paper{}  \def\figlabel{} 
\long\def\nextfig#1{\setcounter{figure}{\value{fignumber}}
  \addtocounter{fignumber}{1}
  \ifnum \viewgraphs=1 \newpage \pagestyle{empty} \fi 
  \ifnum\value{onefig}=0 #1 \fi                 
  \ifnum\value{onefig}=\value{fignumber} #1 \fi}
\def\figwidths#1#2{\ifnum \nocaptions=1 #2mm \else #1mm \fi}  
\def\paper#1{}  
\long\def\plotfig#1#2{\ifnum \nofigures=1 \else #2 \fi}
\long\def\captiontext#1{\ifnum \nofigures=1 \raggedright \fi 
   \ifnum \nocaptions=1 \paper
     \ifnum \viewgraphs=0 
       \newline  \mbox{}\hrulefill\mbox{} \newline 
       \newline label:~\{\figlabel\} 
     \fi 
     \else \ifnum \nofigures=0 \fi 
   #1 \fi}

\newcount\panelwidth \newcount\panelheight 
\newcount\bxmin \newcount\bymin \newcount\bxmax \newcount\bymax
\newcount\tbxmin \newcount\tbymin
\newcount\tpanelwidth \newcount\tpanelheight \newcount\tpdif
\def\panelsize #1,#2;{\panelwidth=#1 \panelheight=#2}  
\def\setbb #1,#2;#3,#4;#5,#6;{
  \tbxmin=#1 \tbymin=#2    
  \bxmin=#3 \bymin=#4      
  \bxmax=#5 \bymax=#6}     
\def\barepanel #1{%
  \ifnum\panelheight=0 
    \tpdif=\bymax \advance\tpdif by -\bymin
    \multiply \tpdif by \panelwidth
    \tpanelheight=\tpdif
    \tpdif=\bxmax \advance\tpdif by -\bxmin
    \divide \tpanelheight by \tpdif
  \else \tpanelheight=\panelheight \fi
  \epsfig{file=#1,%
     bbllx=\bxmin bp,bblly=\bymin bp,bburx=\bxmax bp,bbury=\bymax bp,clip=,%
     width=\panelwidth mm,height=\tpanelheight mm}}
\def\labelypanel #1{
  \ifnum\panelheight=0 
    \tpdif=\bymax \advance\tpdif by -\bymin
    \multiply \tpdif by \panelwidth
    \tpanelheight=\tpdif
    \tpdif=\bxmax \advance\tpdif by -\bxmin
    \divide \tpanelheight by \tpdif
  \else \tpanelheight=\panelheight \fi
  \tpdif=\bxmax \advance\tpdif by -\tbxmin
  \tpanelwidth=\panelwidth \multiply \tpanelwidth by \tpdif
  \tpdif=\bxmax \advance\tpdif by -\bxmin
  \divide \tpanelwidth by \tpdif
  \epsfig{file=#1,%
    bbllx=\tbxmin bp,bblly=\bymin bp,bburx=\bxmax bp,bbury=\bymax bp,%
    clip=,width=\tpanelwidth mm,height=\tpanelheight mm}}
\def\labelxpanel #1{%
  \ifnum\panelheight=0 
    \tpdif=\bymax \advance\tpdif by -\bymin
    \multiply \tpdif by \panelwidth
    \tpanelheight=\tpdif
    \tpdif=\bxmax \advance\tpdif by -\bxmin
    \divide \tpanelheight by \tpdif
  \else \tpanelheight=\panelheight \fi
  \tpdif=\bymax \advance\tpdif by -\tbymin
  \multiply \tpanelheight by \tpdif
  \tpdif=\bymax \advance\tpdif by -\bymin
  \divide \tpanelheight by \tpdif
  \epsfig{file=#1,%
    bbllx=\bxmin bp,bblly=\tbymin bp,bburx=\bxmax bp,bbury=\bymax bp,%
    clip=,width=\panelwidth mm,height=\tpanelheight mm}}
\def\labelxypanel #1{%
  \ifnum\panelheight=0 
    \tpdif=\bymax \advance\tpdif by -\bymin
    \multiply \tpdif by \panelwidth
    \tpanelheight=\tpdif
    \tpdif=\bxmax \advance\tpdif by -\bxmin
    \divide \tpanelheight by \tpdif
  \else \tpanelheight=\panelheight \fi
  \tpdif=\bxmax \advance\tpdif by -\tbxmin
  \tpanelwidth=\panelwidth \multiply \tpanelwidth by \tpdif
  \tpdif=\bxmax \advance\tpdif by -\bxmin
  \divide \tpanelwidth by \tpdif 
  \tpdif=\bymax \advance\tpdif by -\tbymin 
  \multiply \tpanelheight by \tpdif
  \tpdif=\bymax \advance\tpdif by -\bymin
  \divide \tpanelheight by \tpdif
  \epsfig{file=#1,%
    bbllx=\tbxmin bp,bblly=\tbymin bp,bburx=\bxmax bp,bbury=\bymax bp,%
    clip=,width=\tpanelwidth mm,height=\tpanelheight mm}}

\def\CC{\par \vspace*{-2ex} \footnotesize \baselineskip=8pt \begin{verbatim}}

\long\def\startignore #1\stopignore{}   

\def\setlistparams{         
  \topsep=0.7ex                 
  \itemsep=0.7ex                
  \leftmargini=3ex}             
\setlistparams                  

\newcounter{alistindex}       

\newcounter{romenumnr}

\newlength{\minipagewidth}

\newsavebox{\boxcontent}
\newcommand{\ovalhead}[1]{
  \unitlength=1cm
  \sbox{\boxcontent}{\mbox{~~{#1}~~}}
  \begin{center}
    \ifdim\wd\boxcontent>6ex 
    \ifdim\wd\boxcontent<8cm 
    \begin{picture}(8,3) \thicklines     
      \put(4.0,0.8){\oval(8,1.6)} 
      \put(0.0,0.7){\parbox{8cm}{
         \begin{center} \usebox{\boxcontent} \end{center}}}
    \end{picture}
    \else \ifdim\wd\boxcontent<12cm 
    \begin{picture}(12,3) \thicklines     
        \put(6.0,0.8){\oval(12,1.6)} 
        \put(0.0,0.7){\parbox{12cm}{
           \begin{center} \usebox{\boxcontent} \end{center}}}
    \end{picture}
    \else
    \begin{picture}(16,3) \thicklines     
        \put(8.0,0.8){\oval(16,1.6)} 
        \put(0.0,0.7){\parbox{16cm}{
           \begin{center} \usebox{\boxcontent} \end{center}}}
    \end{picture}
    \fi \fi \fi
  \end{center}} 

\newcounter{headnr}            
\newcounter{subheadnr}[headnr]
\newcounter{subsubheadnr}[subheadnr]
\def\head #1\par{
  \stepcounter{headnr}                          
  \vspace{2ex} \noindent                        
  {\bf \theheadnr~~~~#1}\\[1ex] \noindent}      
\def\subhead #1\par{  
  \stepcounter{subheadnr}
  \vspace{1.3ex} \noindent
  {\bf \theheadnr.\arabic{subheadnr}~~~#1}\\[0.3ex] \noindent}
\def\subsubhead #1\par{
  \stepcounter{subsubheadnr}
  \vspace{1.0ex} \noindent
  {\bf \theheadnr.\arabic{subheadnr}.\arabic{subsubheadnr}~~~#1}\\ \noindent}

\font\dropfont= cmr12 scaled \magstep5
\def\dropcap#1#2{{\noindent
    \setbox0\hbox{\dropfont #1}\setbox1\hbox{#2}\setbox2\hbox{(}%
    \count0=\ht0\advance\count0 by\dp0\count1\baselineskip
    \advance\count0 by-\ht1\advance\count0by\ht2
    \dimen1=.5ex\advance\count0by\dimen1\divide\count0 by\count1
    \advance\count0 by1\dimen0\wd0
    \advance\dimen0 by.25em\dimen1=\ht0\advance\dimen1 by-\ht1
    \global\hangindent\dimen0\global\hangafter-\count0
    \hskip-\dimen0\setbox0\hbox to\dimen0{\raise-\dimen1\box0\hss}%
    \dp0=0in\ht0=0in\box0}#2}

\def\level #1 #2#3#4{$#1 \: ^{#2} \mbox{#3} ^{#4}$}   

\def\Teff{\hbox{$\rm{T}_{\rm eff}$}}            

\def\vt{\hbox{$\xi_t$}}                     

\def\mic{\hbox{$\mu$m}}                     

\newcommand{\bi}{\begin{itemize}}
\newcommand{\ei}{\end{itemize}}

\def\mathstacksym#1#2#3#4#5{\def#1{\mathrel{\hbox to 0pt{\lower 
    #5\hbox{#3}\hss} \raise #4\hbox{#2}}}}

\mathstacksym\lta{$<$}{$\sim$}{1.5pt}{3.5pt} 
\mathstacksym\gta{$>$}{$\sim$}{1.5pt}{3.5pt} 
\mathstacksym\lrarrow{$\leftarrow$}{$\rightarrow$}{2pt}{1pt} 
\mathstacksym\lessgreat{$>$}{$<$}{3pt}{3pt} 

\title[Estimating Stellar Parameters from Spectra using a Hierarchical Bayesian Approach]{Estimating Stellar Parameters from Spectra using a Hierarchical Bayesian Approach \thanks{Based on observations with ISO, an ESA
project with instruments funded by ESA Member States (especially the
PI countries France, Germany, the Netherlands and the United Kingdom)
and with the participation of ISAS and NASA.}   }

\author[Z. Shkedy, L. Decin, G. Molenberghs, and
  C. Aerts]{Z. Shkedy$^{1}$\thanks{E-mail: ziv.shkedy@uhasselt.be},
  L. Decin$^{2}$\thanks{{\em Postdoctoral Fellow of the Fund for
  Scientific Research, Flanders}; e-mail:
  Leen.Decin@ster.kuleuven.be}, G. Molenberghs$^{1}$, and
  C. Aerts$^{2}$\\ $^{1}$ Center for Statistics, Hasselt University,
  Agoralaan, B-3590 Diepenbeek, Belgium \\ $^{2}$ Department of
  Physics and Astronomy, Institute for Astronomy, K.U.Leuven,
  Celestijnenlaan 200B, B-3001 Leuven, Belgium}

\begin{document}

\date{Accepted date. Received date; in original form date}

\pagerange{\pageref{firstpage}--\pageref{lastpage}} \pubyear{2006}

\maketitle

\label{firstpage}


\begin{abstract}
A method is developed for fitting theoretically predicted astronomical spectra
to an observed spectrum. Using a hierarchical Bayesian principle, the method
takes both systematic and statistical measurement errors into account, which has
not been done before in the astronomical literature. The goal is to estimate
fundamental stellar parameters and their associated uncertainties.  The
non-availability of a convenient deterministic relation between stellar
parameters and the observed spectrum, combined with the computational
complexities this entails, necessitate the curtailment of the continuous
Bayesian model to a reduced model based on a grid of synthetic spectra. A
criterion for model selection based on the so-called predictive squared error
loss function is proposed, together with a measure for the goodness-of-fit
between observed and synthetic spectra. The proposed method is applied to the
infrared 2.38--2.60\,\mic\ ISO-SWS data (Infrared Space Observatory - Short
Wavelength Spectrometer) of the star $\alpha$ Bootis, yielding estimates for the
stellar parameters: effective temperature \Teff\,=\,4230\,$\pm$\,83\,K, gravity
$\log$ g\,=\,1.50\,$\pm$\,0.15\,dex, and metallicity [Fe/H]\,=\,$-0.30 \pm
0.21$\,dex.
\end{abstract}

\begin{keywords}
Methods: data analysis -- Methods: statistical -- Techniques:
spectroscopic -- Stars: fundamental parameters -- Stars: individual:
Alpha Boo
\end{keywords}


\section{Introduction}
There are two general approaches to the observational study of stellar atmospheres: {\em analysis\/} and {\em synthesis}. Analysis entails measuring detailed features of the spectrum under investigation and hence deducing the parameters of the stellar atmosphere. Synthesis implies specifying atmospheric parameters and calculating the resulting spectrum: when the synthetic and observed spectra agree sufficiently closely and/or in an optimal way, the parameters associated with the synthetic spectrum are taken as estimates for the star under consideration. Current applications of the synthesis technique in the astronomical literature are, however, hampered by the lack of a suitable objective method for deciding which one out of a pool of candidate synthetic spectra matches the observed one best. Often, the observed spectrum is simply presented
along with a ``best'' synthetic spectrum without any mention of the fit criteria employed. Oftentimes, visual comparison is used, which may be adequate if the spectral region used is relatively short and contains only a few spectral lines. Such an eye-fitting method is in danger of failing when the observational data cover a large wavelength range, in which many atomic and/or molecular transitions occur. Moreover, when one wants to account for measurement errors, the task of deciding upon the ``best'' synthetic spectrum is even more complicated.

Inferences for parameters of a stellar atmosphere using the synthesis approach consist of comparing the observed spectrum of the star with a collection of synthetic spectra. Let $\Omega=($\Teff, \mbox{$\log$ g}, \mbox{[Fe/H]}) be the most important parameters of the stellar atmosphere:  temperature in Kelvin, gravity expressed on the log scale, and metallicity. Let $M$ refer to the number of synthetic spectra in the grid. A synthetic spectrum, $\theta^{(m)}$ ($m=1,\dots,M$) is identified  by its value for $\Omega$, $\Omega^{(m)}$ say.

Previously employed frequentist parameter estimation and model selection for the spectrum are based on a goodness-of-fit statistic, $T(y,\theta^{(m)})$, measuring the discrepancy between observed and synthetic spectra. Kolmogorov-Smirnov test statistics and residual sum of squares are discussed in \citet{Decin2000A&A...364..137D,Decin2004A&A...421..281D}.  Both methods use the value of $\Omega^{(m)}$ minimising $T(y,\theta^{(m)})$ as an estimate for $\Omega$. { However, two extra complexities render a paradigm shift a sensible approach, away from frequentist and towards Bayesian methods. Of course, this assertion does not imply the Bayesian paradigm should be deemed in any way superior over the likelihood and/or frequentist paradigms.} First, the analysis presented in Sect.~\ref{application} reaches a very high level of agreement between observed and theoretical data sets. A proper
inclusion of \emph{both} systematic and statistical measurement errors in the model selection and parameter determination procedure is then in its place. Second, the computation of the theoretical data takes many CPU-hours, rendering the calculation of a huge grid of theoretical spectra unfeasible.

Here, we present an objective tool, based on hierarchical Bayesian ideas, for measuring the goodness-of-fit between observational and synthetic spectra, at the same time incorporating the statistical and systematic measurement errors. { Precisely, the reason for choosing the Bayesian paradigm  is the ability to combine the observed spectra with prior knowledge.
Such prior knowledge, termed {\em  expert priors\/}, originates from the theory of and empirical knowledge gathered about stellar atmospheres.} The proposed method is suitable for estimating stellar parameters, other than the ones presented here.  For readers not used to Bayesian statistics, the main principles are outlined in Sect.~\ref{bayesian1}, supplemented with key references.

Sect.~\ref{data} introduces the data setting. A hierarchical Bayesian model is presented in Sect.~\ref{bayesianmodel}, while
the tasks of calculating the prior distribution and model selection issues are discussed in Sects~\ref{posterior} and \ref{modelselection}, respectively.  As in \citet{Decin2004A&A...421..281D}, we apply our method to the case study of the 2.38--2.60\,\mic\ ISO-SWS spectrum of the K2IIIp star Alpha Bootis (Arcturus, HD~124897). Sect.~\ref{application} is devoted to the application. In Sect.~\ref{Discussion}, we compare the results as obtained from the Bayesian methodology with other studies. 


\section{Bayesian inference} \label{bayesian1}
\subsection{Bayes' theorem and marginalisation} \label{bayesian}
{ To support understanding in this and subsequent sections, Table~\ref{tabsymbol} presents the main symbols used.
\begin{table}
\caption{\em Symbols used in the proposed Bayesian method.}
\label{tabsymbol}
\begin{center}
\begin{tabular}{|l|l|}
\hline
\hline
 symbol & meaning\\ \hline
$y$     & observed spectrum \\
$\theta$ & synthetic spectrum \\ 
$\mu$    & ``true'' spectrum              \\ 
$\sigma^{2}_{M}$& SPARE-tag \\ 
$\sigma^{2}$& STDEV-tag \\ 
$\Omega$ &triplet of stellar parameters \\ \hline
$P(y|\mu)$ & likelihood function \\ 
$P(\mu|\theta,\sigma^{2}_{M})$& spectrum's prior distribution\\ 
$P(\mu|y,\theta,\sigma^{2},\sigma^{2}_{M})$& spectrum's posterior distribution\\ 
$T^{m}(y,\mu)$ & goodness-of-fit score for model selection \\ \hline\hline
\end{tabular}
\end{center}
\end{table}
}

Similar to the frequentist inferential approach, the Bayesian paradigm is based on observations, $y$, taken with
uncertainty and assumed to be sampled from a population distributed according to a probability distribution function,
$P(y|\pi)$. While within the frequentist framework a parameter $\pi$ is assumed to be an unknown constant, inference then being based on the sampling distribution of the data given the parameter, i.e., the likelihood function $P(y|\pi)$, the Bayesian approach entertains the idea that $\pi$ is a random variable with a so-called prior distribution, $P(\pi)$ and with inference proceeding based on the conditional distribution of the parameter given the data $P(\pi|y)$, the so-called posterior distribution. The latter follows from the prior distribution and likelihood function combined, using Bayes' theorem (Eq.\ (\ref{bayes})) and the concept of marginalisation (Eq.\ (\ref{marginalisation})):
\begin{equation}
P(\pi|y) = \frac{P(y|\pi) \times P(\pi)}{P(y)}\,,
\label{bayes}
\end{equation}
and
\begin{equation}
P(y) = \int\limits_{-\infty}^{+\infty} P(y|\pi)P(\pi)\, {\rm{d}}\pi\,.
\label{marginalisation}
\end{equation}
The prior probability represents our state of knowledge about the distribution of the parameter before we analyse
the data. This knowledge is modified by the experimental measurements through
the likelihood function, producing the posterior distribution. 
When omitting $P(y)$ from Eq.~(\ref{bayes}), one writes $P(\pi|y) \propto P(y|\pi) \times P(\pi)$.
This is fine for many statistical inferences, such as parameter and precision estimation. However, when model selection is envisaged, the term $P(y)$, often termed {\em evidence\/}, is vitally important.


\subsection{Some examples} \label{examples}

\subsubsection{Example 1} \label{example}
Consider a single observation, $y$, from a normal distribution with
mean $\theta$ and known variance $\sigma^{2}$. The likelihood in this
case is
\[
P(y|\theta)=\frac{1}{\sqrt{2 \pi \sigma}}
\exp \left ( \frac{1}{2 \sigma^{2}} (y-\theta)^{2} \right )
\propto
\exp \left ( \frac{1}{2 \sigma^{2}} (y-\theta)^{2} \right ).
\]
Assuming further that  $\theta$ is normally distributed with mean
$\mu$ and variance $\tau^{2}$, the prior model is
\[
P(\theta|\mu,\tau) \propto
\exp \left ( \frac{1}{2 \tau^{2}} (\theta-\mu)^{2} \right ).
\]
\citet{Gelman1995} derived the posterior distribution to be
\begin{equation}
P(\theta|y) \propto
\exp \left ( \frac{1}{2 \delta^{2}} (\theta-\eta)^{2} \right ),
\label{normalposterior2}
\end{equation}
which is a normal distribution with mean $\eta$ and
 variance $\delta^{2}$.
We return to the parametric structure of $\eta$ and
$\delta^{2}$ in Sect.~\ref{sectspecific}.

\subsubsection{Example 2} \label{example2}

Consider a sequence of $n$ Bernoulli, i.e., 0/1, trials $y_{1},\dots,y_{n}$,  with probability of observing 1 equal to $\theta$, and let $y=\Sigma_{i=1}^{n}y_{i}$.  
The resulting binomial likelihood is given by
\[
P(y|\theta) \propto \theta^{y}(1-\theta)^{(n-y)},
\]
and the (\emph{frequentist}) maximum likelihood for the success
probability $\theta$ is 
$\theta_{ML}=y/n$. Suppose we specify the prior distribution for the success
probability to be Beta: $\theta \sim
\mbox{Beta}(\alpha,\beta)$, then
\[
P(\theta) \propto \theta^{\alpha-1}(1-\theta)^{\beta-1}.
\]
Then, the prior mean for $\theta$ is $\alpha/(\alpha+\beta)$. The posterior
distribution of $\theta$ then is:
\[
P(\theta|y) \propto \theta^{(y+\alpha-1)}(1-\theta)^{(n-y+\beta-1)},
\]
which is, again, a beta distribution, $\theta|y\sim
\mbox{Beta}(\alpha+y,\beta+n-y)$, with posterior mean 
\[
E(\theta|y)=\frac{\alpha+y}{\alpha+\beta+n}.
\]
To illustrate this model further, assume 6 successes were obtained out of 10 trials and suppose that we
specify a non-informative prior $\theta \sim U(0,1)$ (hence,
$\alpha=\beta=1$ since $\mbox{Beta}(1,1)$ is a uniform distribution over the
interval $[0,1]$). In this case, the maximum likelihood estimate for $\theta$ is
$\widehat{\theta}=0.6$ when using the classical
frequentist methodology, while the Bayesian analysis results in a
 posterior mean of $\widehat{\theta}=7/12=0.583$. In other words, since the posterior mean is obtained by pulling the maximum likelihood value $0.6$ towards the prior mean of the $U(0,1)$, which equals 5. The larger the sample size, the less the importance of the prior distribution.


\subsubsection{Hierarchical models} \label{hierarchical}

The above examples can be formulated as \emph{hierarchical models}
in which the likelihood and the prior are specified at the first
and the second level of the model. At the third level of the model
we specify the probability model for the \emph{hyperparameters}
$\tau^{2}$ and  $\sigma^{2}$, $F_{\tau}$ and $F_{\sigma}$, which
are called \emph{hyperprior distributions}. Hence, we obtain e.g.\ for
example 1:
\[
 \begin{array}{ll}
                 y \sim N(\theta,\sigma^{2}),&\;\;\;\mbox{1st level}, \\
                 \theta \sim N(\mu,\tau^{2}),&\;\;\;\mbox{2nd level},\\
                 \tau^{2} \sim F_{\tau}\;\mbox{and}\; \sigma^{2} \sim F_{\sigma},&\;\;\;\mbox{3rd level}.\\
                    \end{array}
\]

\subsection{Posterior inference} \label{posterior_example}

As was explained in Sect.~\ref{bayesian}, inference is based on the posterior distribution of the unknown parameters in the model given the data $P(\pi|y)$.  This distribution can be derived analytically (as in the above example) or may have to be approximated using the so-called Markov Chain Monte Carlo (MCMC) algorithm. A single iteration of the MCMC algorithm \citep{Gilks1996} consists of sampling the unknown parameters in the model from their full conditional distribution, given the current value of the other parameters in the model and the data. Assume that the distribution of interest is $P(\mu)$, where $\mu=(\mu_{1},\dots,\mu_{d})$. We denote the full conditional distribution of $\mu_{i}$ given all other parameters by $P(\mu_{i}|\mu_{-i})$.

One way to implement the MCMC algorithm is through the well-known Gibbs sampling algorithm \citep{Gilks1996}, the steps of which are as follows:
\begin{itemize}
\item Step 1:\newline Initialize the iteration counter of the chain
($j=1$) and the initial values for the parameters
$\mu^{(0)}=(\mu^{(0)}_{1},\dots,\mu^{(0)}_{d})$.
\item Step 2: \newline Draw a new value
$\mu^{j}=(\mu^{(j)}_{1},\dots,\mu^{(j)}_{d})$ through successive
sampling from the full conditional distributions:
\[
 \begin{array}{ll}
\mu_{1}^{(j)} \sim P(\mu_{1}|\mu_{2}^{(j-1)},\dots,\mu_{d}^{(j-1)}),\\
\mu_{2}^{(j)} \sim P(\mu_{2}|\mu_{1}^{(j)},\mu_{3}^{(j-1)}\dots,\mu_{d}^{(j-1)}),\\
. \\
\mu_{i}^{(j)} \sim P(\mu_{i}|\mu_{1}^{(j)},\dots,\mu_{i-1}^{(j)},\mu_{i+1}^{(j-1)}\dots,\mu_{d}^{(j-1)}),\\
. \\
\mu_{d}^{(j)} \sim P(\mu_{d}|\mu_{1}^{(j)},\dots,\mu_{d-1}^{(j)}).\\
\end{array}
\]
\item
Repeat the second step until convergence.
\end{itemize}
Assuming that the sampling process is converged after $L$
iterations, the posterior mean of $\mu$ can be estimated by MCMC integration:
\[
\bar{\mu}_{i}=\sum_{\ell=1}^{L}\frac{\mu_{i}^{(\ell)}}{L}.
\]
Note that  $\bar{\mu}_{i}$ is simply the sample mean of $\mu_{i}$
which is obtained after $L$ iterations of the Gibbs sampling. In our
setting $\bar{\mu}_{i}$ is the posterior mean of the spectrum at
wavelength $i$.

One of the quantities of interest will be $T^{(m)}(y,\mu^{\ell})$ (see
Sect.~\ref{goodness}).
In practice, if we draw $L$ simulations from the posterior
distribution of $\mu$ we can monitor the value of
$T^{(m)}(y,\mu^{\ell})$ for each iteration, $\ell=1,2,\ldots,L$ and the
posterior mean of $T^{(m)}(y,\mu^{\ell})$ is simply $1/L \sum_{\ell=1}^{L}
T^{(m)}_{\ell}(y,\mu)$.

{ It is important to realise that the Bayesian method is typically based on fully specifying the likelihood function, together with a prior distribution. These, combined with the data, produce the posterior distribution and ultimately statistical inferences. When analytic computations are deemed too cumbersome, one may then elect MCMC computations instead. Such a switch does not change the parametric nature of the assumptions made and hence the MCMC implementation is fully parametric. Furthermore, in many instances, like the one considered here, opting for normal distributions greatly simplifies computations.}

\section{Observed and synthetic spectra} \label{data}
Let us discuss the observational data setting and the concept of
synthetic spectra in turn.

\subsection{Observational data $\textbfit{y}$} \label{observations}
The observational data for $\alpha$ Boo, also considered in
\citet{Decin2004A&A...421..281D} consist of near-infrared
(2.38--2.60\,\mic, band 1A) spectra, observed with the SWS
\citep[Short Wavelength Spectrometer,][]{deGraauw1996A&A...315L..49D}
on board ISO \citep[Infrared Space
Observatory,][]{Kessler1996A&A...315L..27K}. Prior to the statistical
analysis, data reduction techniques are applied
\citep{Decin2004A&A...421..281D}.  {\em Bands\/} are combinations of
detector array, aperture and grating orders such that for each band
its detector array sees a unique order of light, and hence a unique
wavelength $\lambda$. Band 1 (2.38--4.08\,\mic) is subdivided in 4
sub-bands: band 1A: 2.38--2.60\,\mic, band 1B: 2.60--3.02\,\mic, band
1D: 3.02--3.52\,\mic, and band 1E: 3.52--4.08\,\mic. The same
resolution and factor shift for band 1A are used as in Table~1 of
\citet{Decin2004A&A...421..281D}. Let us turn to the uncertainties and
errors in the data.

The error propagation of the SWS pipeline separates statistical errors
from systematic ones. The so-called statistical `STDEV-tag' $\sigma$
contains the standard deviation of the points in a given bin, and the
systematic `SPARE-tag' $\sigma_M$ captures the effect of the imperfect
performance of the instrument. The `SPARE-tag' of the ISO-SWS data
corresponds to statistical accuracy, i.e., how well systematic errors
can be controlled, closeness between the result of an experiment and
the true value, while the `STDEV-tag' corresponds to the precision,
i.e., how well the random errors can be controlled. The errors
$\sigma$ and $\sigma_M$ have the same order of magnitude
(Fig.~\ref{fig1}).

While the `SPARE-tags' are almost the same for all observations of all
target stars observed by the satellite, the `STDEV-tag' discriminates
between the quality of the data points.  Assume a normally distributed
model for the observed spectrum $y_i$ at wavelength $i$,
$i=1,2,\dots,n$ with mean ${\rm{E}}(y_i) = \mu_i$, representing the
true spectrum of the target, possibly including systematic
instrumental artifacts, and statistical measurement error variance
$\sigma_i$, then a normal model
\begin{equation}
\label{like0}
y_{i}=\mu_{i}+\varepsilon_{i},
\end{equation}
where $\varepsilon_{i}\sim N(0,\sigma^{2}_{i})$ is assumed.

\subsection{Synthetic data $\btheta$}\label{syntdata}
A synthetic stellar spectrum is computed from first-principle physics
laws governing the stellar atmosphere.  For a full description of this
study's synthetic spectra we refer to \citet{Decin2000A&A...364..137D,
Decin2004A&A...421..281D}. It is very important to note that the
functions of interest are of a continuous nature, yet they will be
treated in a discretized way, for reasons of numerical
feasibility. Indeed, the synthetic spectra  calculations require a
model atmosphere as input, which is obtained through lengthy
calculations, taking several hours, in order to obtain hydrostatic
equilibrium and to fulfill the conservation law of radiative (and
convective) energy. When this would not have been the case, i.e., when
we could have written $\mu(\lambda) = h($\Teff$,\log {\rm{g}},
{\rm{[Fe/H]}}, \lambda)$, with $h$ representing a closed analytical
function, then we could have estimated \Teff, $\log$ g and [Fe/H]
\emph{directly} from the observational spectrum.  We circumvent the
absence of a closed form for the spectrum by considering a dense grid
of synthetic spectra $\theta$, with the goal of providing appropriate
error estimates.

Subsequently, we rely on hierarchical Bayesian models for spectrum
fitting, following the idea proposed by \citet{Laud1995} and
\citet{Gelfand1998}, who suggested comparing the observed data
($y$) and hypothetical data, termed \emph{replicated data}, sampled
from the posterior predictive distribution, by minimising a predictive
discrepancy measure (Sect.~\ref{modelselection}).

\section{Hierarchical Bayesian model for the spectrum}\label{bayesianmodel}
Applying Bayes' theorem produces the spectrum's posterior
distribution, as outlined in Sect.~\ref{bayesian1}.
Precisely, from our knowledge of $\theta$ and $y$, we predict $\mu$, i.e., we derive its
posterior. For a ``bad'' synthetic spectrum $\theta$, the
observational data $y$ and the predicted $\mu$ will differ by a
relatively large amount.

Using (\ref{like0}), the likelihood of the model parameters given the
data equals
\begin{equation}
\label{like1}
P(y|\mu,\sigma^{2}) = \prod_{i=1}^{n}\phi(y_i|\mu_{i},\sigma^{2}_{i})\,,
\end{equation}
where $\phi$ is the density of the normal distribution
with parameters $\mu$ and $\sigma^2$.
Assume that the mean of the observational data at wavelength $i$,
$\mu_i$, follows a normal distribution, i.e.,
\begin{equation}
\mu_{i}=\theta_{i}+u_{i},
\label{mumodel}
\end{equation}
with $u_{i} \sim N(0,\sigma^{2}_{M_i})$ and $\sigma_{M_i}$ the
systematic observational error.  Following (\ref{mumodel}), we assume
that, owing to the systematic errors, the true spectrum is distributed
around $\theta_{i}$ with variance $\sigma_{M_i}$.  It follows from
(\ref{mumodel}) that the prior distribution is given by
\begin{equation}
\label{prior1}
P(\mu|\theta,\sigma^{2}_{M})= \prod_{i=1}^{n}\phi(\mu|\theta_{i},\sigma^{2}_{M_i}).
\end{equation}
Then, the spectrum's posterior distribution is
\begin{eqnarray}
P(\mu|y,\sigma^2,\sigma_M^2,\theta) 
& \propto&
P(y|\mu,\sigma^2) \cdot P(\mu|\theta,\sigma_M^2)
\nonumber \\
 & = & \prod_{i=1}^{n}\phi(y_i|\mu_{i},\sigma^{2}_{i})
\cdot \prod_{i=1}^{n}\phi(\mu|\theta_{i},\sigma^{2}_{M_i}).
\label{post1}
\end{eqnarray}

\section{Posterior distribution for the spectrum}\label{posterior}

\subsection{The full model\label{sectfull}}
The above specifications are sufficient to define the posterior
distribution of all model parameters jointly:
\begin{eqnarray}
\lefteqn{P(\mu,\theta,\sigma^2,\sigma_M^2,\Omega|y)} \nonumber \\
&  \stackrel{(\ref{bayes}),(\ref{marginalisation})}{\propto} &
\underbrace{P(y|\mu,\sigma^2)}_{\rm{likelihood,\
Eq.\ (\ref{like1})}} \times
\underbrace{P(\mu|\theta,\sigma_M^2)}_{\rm{prior,\ Eq.\
(\ref{prior1})}} \nonumber \\
& & \times
\underbrace{P(\theta|\Omega)}_{\rm{distribution\ of\ the\ prior\ mean\ }
\theta} \times
\underbrace{P(\Omega)}_{\rm{hyperprior}}.
\label{fullmodel}
\end{eqnarray}
We still need to specify the hyperpriors for $P(\Teff)$, $P(\log g)$,
 and $P(\mbox{[Fe/H]})$.  A literature study for the stellar
 atmosphere parameters of $\alpha$ Boo was presented in \citet{Decin2000A&A...364..137D}, who found that \Teff\ ranges from $4060$\,K to
 $4628$\,K, $\log$ g from $0.90$ to $2.60$\,dex, and
 $[\mbox{Fe}/\mbox{H}]$ from $-0.77$ to $0.00$\,dex, based on which we
 construct the hyperprior distributions.  Further discussion on the
 choice of the grid parameters and the uncertainties thereon is relegated to Sect.~\ref{application}.

After establishing $P(\Omega)$, $P(\theta|\Omega)$ is needed to
complete the specification of the hierarchical model.  Since there is
no deterministic relationship between $\theta$ and $\Omega$, we cannot
specify the mean of the prior distribution using standard methods, including linear,
generalised linear, or non-linear models, for example. This
implies the need to adopt a two-stage approach with calculation of a
collection of models for the synthetic spectrum over a grid of
discrete values in $\Omega$, $\Omega^{(1)},\dots,\Omega^{(M)}$,
followed by usage of these models $\theta^{(m)}$ ($m=1,\dots,M$), as
the prior mean of $\mu$ in (\ref{prior1}). In this approach, the value
of $\Omega$, given the data, is not estimated with the posterior means
of the hyperprior distributions, but rather we select models from the
collection calculated in the first stage. Thus, our two-stage approach
implies a model selection procedure ought to be used to select the
`best' synthetic spectrum. This issue is discussed further in
Sect.~\ref{modelselection}.

\subsection{The reduced model} \label{sectreduced}
For the $m$th combination of $\Omega$, we calculate $\theta^{(m)}$ and
consider a reduced posterior distribution
\begin{eqnarray}
P(\mu,\theta^{(m)},\sigma^2,\sigma_M^2|y)
& \propto & P(y|\mu,\sigma^{2},\theta^{(m)}) \cdot
P(\mu|\theta^{(m)},\sigma^{2}_{M}) \nonumber\\ 
& \propto &
P(\mu|y,\sigma^{2},\sigma^{2}_{M},\theta^{(m)}), \label{reduce11}
\end{eqnarray}
where $P(\mu|y,\sigma^{2},\sigma^{2}_{M},\theta^{(m)})$ is the
posterior distribution of the spectrum $\mu$ given $\Omega^{(m)}$, and
$\theta^{(m)}$ is the prior mean of $\mu$ as in (\ref{prior1}). Since,
for the $m$th combination
$P(\theta^{(m)}|\Omega^{(m)},m)=P(\Omega^{(m)}|m)=1$, passing from
(\ref{fullmodel}) to (\ref{reduce11}) is straightforward.

\subsubsection{Specification of the reduced model} \label{sectspecific}
We focus on the posterior distribution of the spectrum $\mu$ at
wavelength $i$ given $y_{i}$, $\theta_{i}^{(m)}$, $\sigma_{i}$, and
$\sigma_{M_i}$. For the remainder of this section we drop superscript
$m$ and subscript $i$.  Since the prior in (\ref{prior1}) is conjugate
to the normal likelihood in (\ref{like1}), the posterior distribution
of the spectrum is normal as well.  Formally, the likelihood and the
prior can be expressed by
\begin{equation}
P(y|\mu,\sigma^{2}) \propto \exp\left (\
-\frac{1}{2\sigma^{2}}(y-\mu)^{2} \right ),
\label{reduce2}
\end{equation}
and
\begin{equation}
P(\mu|\theta,\sigma^{2}_{M}) \propto \exp\left (\
-\frac{1}{2\sigma^{2}_{M}}(\mu-\theta)^{2} \right ),
\label{reduce3}
\end{equation}
respectively.
It follows from (\ref{reduce2}) and  (\ref{reduce3}) that the posterior
distribution of $\mu$ is
\begin{equation}
\label{reduce4a}
P(\mu|y,\theta,\sigma^{2},\sigma^{2}_{M}) \propto \exp \left ( \
-\frac{1}{2\delta^{2}}(\mu-\theta_{1})^{2}  \right ),
\end{equation}
which is a normal distribution with mean $\theta_1$ and variance
$\delta^2$ given by
\begin{equation}
\label{reduce5}
\theta_{1}=\frac{\frac{1}{\sigma_{M}^{2}}\theta+\frac{1}{\sigma^{2}}y}{\frac{1}{\sigma_{M}^{2}}+\frac{1}{\sigma^{2}}}\;\;\;\;\mbox{and}\;\;\;\;
\frac{1}{\delta^{2}}=\frac{1}{\sigma^{2}}+\frac{1}{\sigma_{M}^{2}}.
\end{equation}
This model is discussed in detail by \citet{Gelman1995}. The
result in (\ref{reduce5}) means that the posterior mean of the
spectrum $\theta_1$ in (\ref{reduce4a}) is a weighted average of the
synthetic spectrum and the observed spectrum. It can be shown that
\begin{equation}
\label{postmean1}
\theta_{1}=\theta+(y-\theta)\cdot
\frac{\sigma_{M}^{2}}{\sigma^{2}+\sigma_{M}^{2}},
\end{equation}
where the second factor on the right hand side is a shrinkage factor.
Hence, if the SPARE-tag $\sigma_{M_i}$, containing the systematic
measurement error, is relatively large compared to the STDEV-tag,
$\sigma_i$, containing the statistical measurement error, the
posterior mean of the spectrum at wavelength $i$ shrinks towards the
observed spectrum at wavelength $i$.  In the reverse case, the
posterior mean of the spectrum shrinks towards the synthetic spectrum.

\subsubsection{Contracting the variance function} \label{variance}
Clearly, the variance parameters $\sigma^{2}$ and $\sigma^{2}_{M_i}$
are unknown and need to be estimated.  Fig.~\ref{fig1} displays the
measurement errors in band 1A (on the log scale). The shrinkage ratio,
$\sigma^2_M/(\sigma^2+\sigma^2_M)$, is shown in panel \textit{c}.
Note that for wavelengths smaller than or equal to 2.4\,$\mu$m the
mean of the shrinkage ratio is 0.5 while for wavelengths greater than
or equal to 2.58\,$\mu$m the mean of the shrinkage ratio increases to
0.87. This means that at the beginning of band 1A the posterior mean
is an average between the observed and synthetic spectrum, while the
weight of the observed spectrum increases with the wavelength.
\begin{figure}
\begin{center}
\resizebox{0.5\textwidth}{!}{\rotatebox{0}{\includegraphics{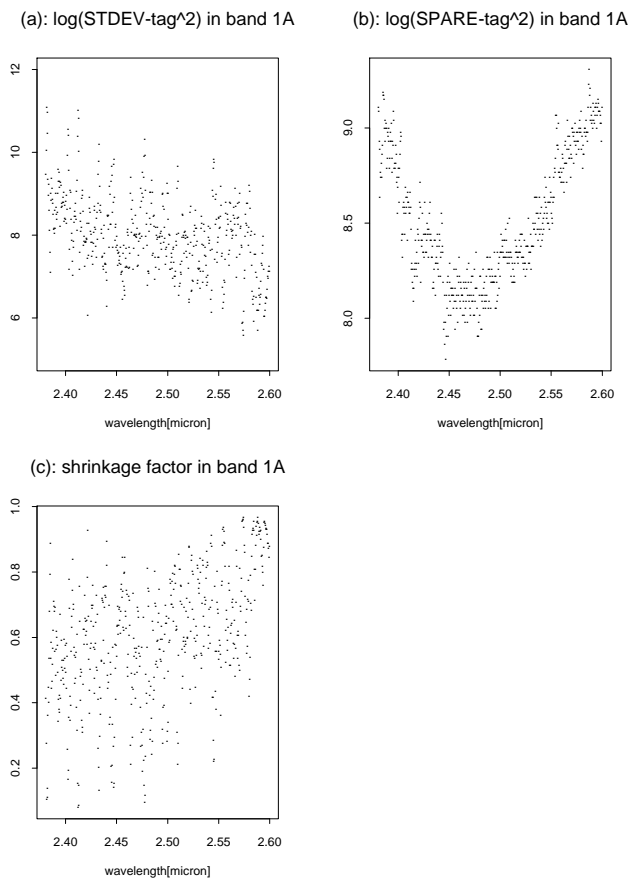}}}
\caption{\em Measurement errors in band 1A. Panel a:
\emph{statistical} $\log$(STDEV-tag)$^{2}$ $(\log(\sigma^{2}))$. Panel b:
\emph{systematic} $\log$(SPARE-tag)$^{2}$ $(\log(\sigma_M^{2}))$. Panel c: the
shrinkage ratio  $\sigma^2_M/(\sigma^2+\sigma^2_M)$.}
\label{fig1}
\end{center}
\end{figure}

To model the variance components, we consider an empirical Bayesian
approach \citep{Carlin1996}. Using the estimates for the measurement
error, we first specify a model for $\sigma^{2}$ and
$\sigma^{2}_{M_i}$, estimate the parameters, and then plug in
predicted values into the model. Specifically, we smooth the data
using a hierarchical linear mixed model \citep{Verbeke2000}, allowing
to estimate a smooth function for the variance components in a
{ flexible} fashion. For $\sigma$, we assume
\begin{equation}
\label{lmm1}
\log(\sigma_{i}^{2}) \sim N(X_{i}\upsilon+Z_{i}u, \delta_{\sigma}^2),
\end{equation}
where $X_{i}$ and $Z_{i}$ are known design matrices, $\upsilon$ are
regression coefficients, and $u = (u_1,\dots,u_K)$ are random effects
assumed to follow $u_{k} \sim N(0,\delta_{u}^2)$ ($k=1,\dots,K$). A
similar model was assumed for $\sigma_M$. Details can be found in
\citet{Zivthesis}.  In this approach, we use the estimated smooth
functions as variance components of the reduced model.  Such a smooth
function allows the data to dominate the posterior mean at the end of
band 1A, where $\sigma$ is relatively small relative to $\sigma_M$.
The application to $\alpha$ Boo is presented in
Sect.~\ref{application}.


\section{Model selection}\label{modelselection}
\subsection{Measures for the goodness-of-fit}\label{goodness}
Using (\ref{reduce11}), we predict $\mu$ from our knowledge on $y$ and
 $\theta^{(m)}$.  Following \citet{Gelman1995} and \citet{Carlin1996},
 a weighted $\chi^{2}$ goodness-of-fit measure, given by
\begin{equation}
\label{chisq1}
T^{(m)}(y,\mu)=\sum_{i=1}^{n}\frac{[y_{i}-E(y_{i}|\mu_{i},\theta^{(m)}]^{2}}{\mbox{var}(y_{i}|\mu_{i},\theta^{(m)})},
\end{equation}
can be used.  $T^{(m)}(y,\mu)$ measures the discrepancy between the
observed data $y$ and the expected mean, relative to the variability
in the model. Both $\sigma_M$ and $\sigma$ influence $T^{(m)}(y,\mu)$,
since $\sigma^2=\mbox{var}(y|\mu)$, and because $\mu \sim
N(\theta^{(m)},\sigma^{2}_{M})$, the denominator depends on both
quantities.

In our application, we will compare the performance of
$T^{(m)}(y,\mu)$ with the results in \citet{Decin2000A&A...364..137D}, who
used a frequentist version of (\ref{chisq1}) that is unable to take
the observational errors into account.  Note, however, that within the
Bayesian framework $T^{(m)}(y,\mu)$ is not used as a criterion for
model selection but rather as a measure for the model goodness-of-fit.

\subsection{Posterior predictive distribution} \label{preddistr}
Criteria for Bayesian model selection are discussed in \citet{Laud1995} and
\citet{Gelfand1998}, all based on the
posterior predictive distribution.

Let $y_{i}$ be the observed data at wavelength $i$ and
$\mu_{i}^{\ell}$ the current value of $\mu_{i}$ at the $\ell$th MCMC
iteration. Then, we simulate $n$ hypothetical replications from the
data given the current value of $\mu_{i}^{\ell}$ and denote these
values by $y^{rep}_{i}$ ($i=1,2,\dots,n$). {F}rom these $n$ replicates
$P(y^{rep}|\mu,\theta,y)$ is constructed.  Formally, the posterior
predictive distribution is given by
\begin{eqnarray}
P(y^{rep}|y) & \stackrel{(\ref{marginalisation})}{=} & \int
P(y^{rep},\mu,\theta)\,d\mu\,d\theta \nonumber \\
 & = &
\int P(y^{rep}|\mu,\theta,y )P(\mu,\theta|y)\,d\mu\,d\theta.
\label{predic1}
\end{eqnarray}
For each replicated sample, obtained from (\ref{predic1}), 
the observed data and the posterior predictive distribution are compared. If
the $m$th synthetic spectrum is sufficiently accurate, the hypothetical
replication and the observed data are considered sufficiently similar.

\subsubsection{Predictive model selection under squared error loss}
A good model for the synthetic spectrum, among the models under
consideration, should render a prediction close to what has been
observed. Thus, a synthetic spectrum model leading to a small
discrepancy between the replication and the observed data is
considered a viable description of the data. A measure for the discrepancy, based on
squared error loss is proposed by \citet{Laud1995}:
\begin{equation}
L_{m}^{2}  =  E[(y^{rep} - y)^T (y^{rep} - y)] 
 =  E\sum_{i=1}^{n}(y^{rep}_{i}-y_{i})^{2}, \label{exploss1}
\end{equation}
where a superscript $T$ refers to transpose.
\citet{Laud1995} and \citet{Gelfand1998} showed that
$L_{m}^{2}$ can be expressed  as a sum of two terms:
\begin{equation}
L_{m}^{2}  = 
 \sum_{i=1}^{n}[E(y^{rep}_{i}-y_{i})^{2}+\mbox{var}(y^{rep}_{i})]
  =  G(m)+P(m). \label{exploss1aa} 
\end{equation}
Here, $G(m)$ measures the goodness-of-fit and $P(m)$ is a penalty
measuring model complexity. The latter is the same for all synthetic
spectra as they are calculated with the same number of parameters.
$L_{m}^{2}$ can now be used for model selection. \citet{Laud1995} and
\citet{Gelfand1998} suggested selecting a model from a
collection of $M$ candidates by minimising the expected squared error
loss of the replicated data.  Hence, the procedure proposed by
\citet{Gelfand1998} requires calculation of $L_{m}^{2}$ over the model
collection:
\begin{equation}
\label{exploss1b}
L_{m}^{2}=\sum_{i=1}^{n}(\eta_{i}^{(m)}-y_{i})^{2}+\sum_{i=1}^{n}\sigma_{i}^{2(m)},
\end{equation}
where $\sigma_{i}^{2(m)}=\mbox{var}(y^{rep}_{i}|y,m)$ and
$\eta_{i}^{(m)}=E(y^{rep}_{i}|y,m)$. In our setting,
$\eta_{i}^{(m)}=E(y^{rep}_{i}|y,\theta^{(m)})$.  If we assume that
both $\sigma_{i}$ and $\sigma_{M_i}$ are known, then the model
minimising $G(m)$ is selected, otherwise the model that minimises
$L_{m}^{2}$ is selected.

{ A schematic representation of the various model building and selection steps is presented in Figure~\ref{flowchart}.}
\setlength{\tabcolsep}{0mm}
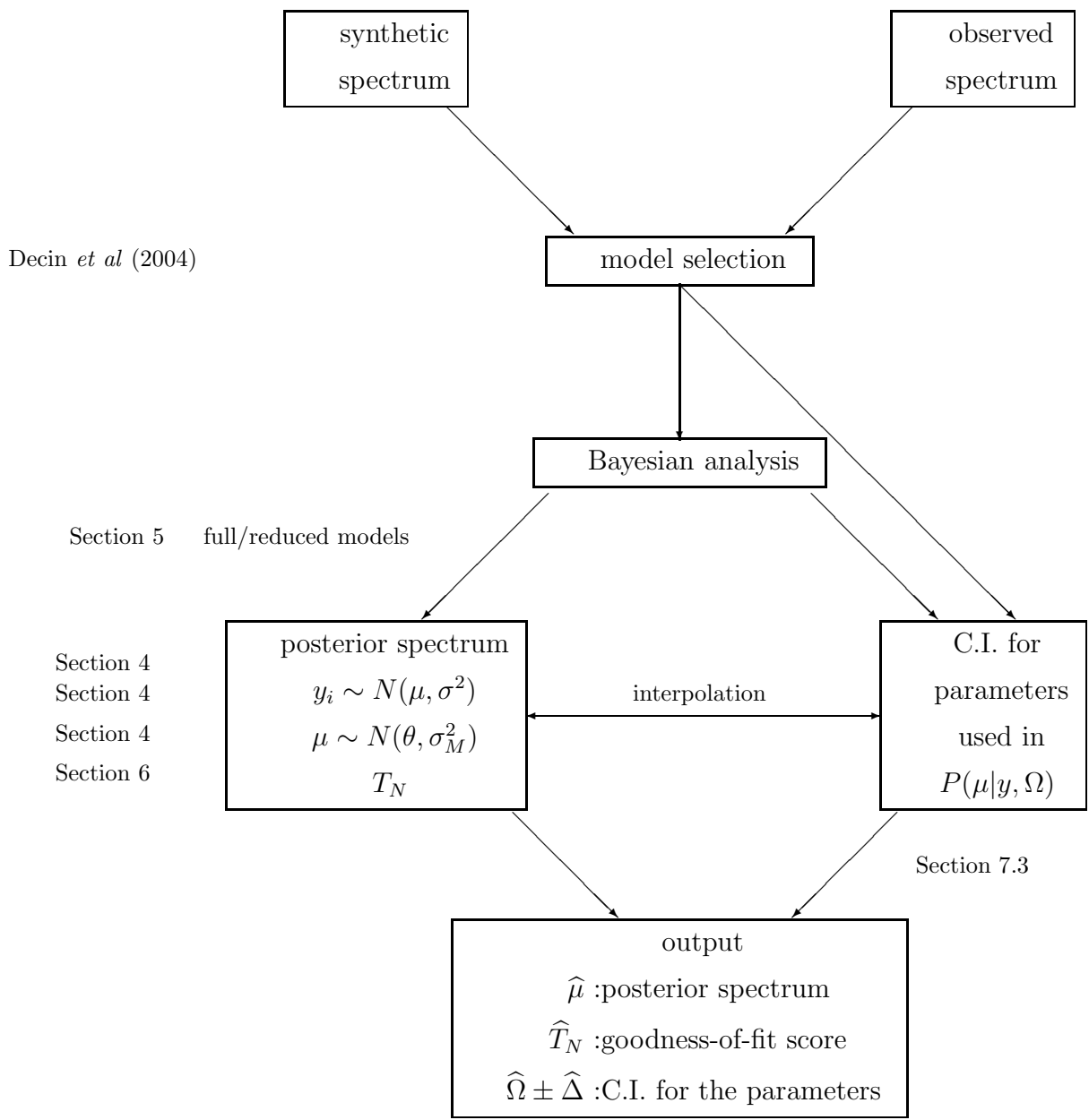
\begin{figure*}
\begin{center}
\setlength{\unitlength}{.15mm}
\begin{picture}(1100,1150)(0,0)

\put(300,1100){\makebox(0,0){\Large\fbox{\begin{tabular}{c} synthetic~~\\[2mm] spectrum~~\end{tabular}}}}
\put(900,1100){\makebox(0,0){\Large\fbox{\begin{tabular}{c} observed~~\\[2mm] spectrum~~\end{tabular}}}}

\put(370,1052){\vector(1,-1){125}}
\put(830,1052){\vector(-1,-1){125}}

\put(10,900){\makebox(0,0){\large {\begin{tabular}{c} Decin {\em et al\/} (2004)  \end{tabular}}}}
\put(600,900){\makebox(0,0){\Large \Large \fbox{\begin{tabular}{c}  model selection~~ \end{tabular}}}}

\put(600,877){\vector(0,-1){155}}

\put(600,877){\vector(1,-1){332}}

\put(10,500){\makebox(0,0){\large {\begin{tabular}{c} Section 4  \end{tabular}}}}
\put(600,700){\makebox(0,0){\Large \Large \fbox{\begin{tabular}{c} Bayesian analysis~~ \end{tabular}}}}

\put(470,670){\vector(-1,-1){125}}
\put(730,670){\vector(1,-1){125}}

\put(300,450){\makebox(0,0){\Large \Large \fbox{\begin{tabular}{c} posterior spectrum~~\\[2mm] $y_i\sim N(\mu,\sigma^2)$~~\\[2mm] $\mu\sim N(\theta,\sigma^2_M)$~~\\[2mm] $T_N$~~ \end{tabular}}}}

\put(900,450){\makebox(0,0){\Large \Large \fbox{\begin{tabular}{c} C.I.\ for~~ \\[2mm] parameters~~\\[2mm] used in~~\\[2mm] $P(\mu|y,\Omega)$~~ \end{tabular}}}}

\put(145,625){\makebox(0,0){\large {\begin{tabular}{c} 
Section 5 $\quad$ full/reduced models
  \end{tabular}}}}

\put(10,450){\makebox(0,0){\large {\begin{tabular}{c} 
\phantom{Tram 10}\\[2mm]
Section 4\\[2mm]
Section 4\\[2mm]
Section 6
  \end{tabular}}}}

\put(600,470){\makebox(0,0){\large {\begin{tabular}{c} 
interpolation  \end{tabular}}}}

\put(448,450){\vector(1,0){351}}

\put(799,450){\vector(-1,0){351}}

\put(435,354){\vector(1,-1){104}}
\put(815,354){\vector(-1,-1){104}}

\put(870,300){\makebox(0,0){\large {\begin{tabular}{c} Section 7.3  \end{tabular}}}}

\put(600,150){\makebox(0,0){\Large \Large \fbox{\begin{tabular}{rcl} 
\multicolumn{3}{c}{output}~\\[2mm]
$\widehat{\mu}$&~:~&posterior spectrum~~\\[2mm]
$\widehat{T}_N$&~:~&goodness-of-fit score~~\\[2mm]
$\widehat{\Omega}\pm \widehat{\Delta}$&~:~&C.I.~for the parameters~~
 \end{tabular}}}}

\end{picture}

\caption{\label{flowchart} \em Schematic representation of the model building and selection steps' sequencing.}
\end{center}
\end{figure*}

\section{Application: The case study of $\balpha$ Boo}\label{application}
We apply the Bayesian method as developed above to the case study of
the 2.38--2.60\,$\mu$m ISO-SWS spectrum of the metal-deficient K2III
peculiar giant $\alpha$ Boo and compare the newly obtained results
with other frequentist studies, in particular with the results of
\citet{Decin2004A&A...421..281D}.  The same set of synthetic spectra
has been used by these authors, i.e., a grid over discrete values in
$\Omega = ($\Teff$, \log {\rm{g}}, {\rm{[Fe/H]}})$, with parameter
values \citep{Decin2000A&A...364..137D}:
\begin{center}
\begin{tabular}{lcl}
\Teff\ & : & 4160\,K, 4230\,K, 4300\,K, 4370\,K, 4440\,K\\
$\log$ g & : & 1.20, 1.35, 1.50, 1.65, 1.80 \\
{[Fe/H]} & : & $0.00$, $-0.15$, $-0.30$, $-0.50$, $-0.70$.
\end{tabular}
\end{center}
As in \citet{Decin2004A&A...421..281D}, other parameters needed to
  compute a proper spherically symmetric atmosphere model and
  synthetic spectrum were kept fixed: the abundance of carbon
  $\varepsilon$(C)\,=\,$7.96 \pm 0.20$, nitrogen
  $\varepsilon$(N)\,=\,$7.61 \pm 0.25$, and oxygen
  $\varepsilon$(O)\,=\,$8.68 \pm 0.20$, and the microturbulent
  velocity \vt\,=\,$1.7 \pm 0.5$\,km/s).  Each synthetic spectrum is
  used as a prior mean in the hierarchical model of
  Sect.~\ref{sectreduced}. There are 125 models in total, labelled by
  an (arbitrary) model number, as listed in Table~\ref{modelnumber}. A
  proper angular diameter was calculated for each model in the grid
  using Eq.~(1) in \citet{Decin2004A&A...421..281D}. The derived
  values are listed in Table~\ref{modelnumber}.
  \addtolength{\tabcolsep}{1mm}
\begin{table*}
\caption{\em \label{modelnumber}Steller angular diameters (expressed
  in milli-arcseconds) and model numbers (in between brackets)
  associated with the different model parameters of the grid of
  synthetic spectra.}
\begin{center}
\setlength{\tabcolsep}{1.2mm}
\begin{tabular}{cclrlrlrlrlrcl}\hline\hline
 && \multicolumn{10}{c}{\Teff\ [K]}
& \\
&\cline{2-11}
{$\log$ g} && \multicolumn{2}{c}{4160} &  \multicolumn{2}{c}{4230} &  \multicolumn{2}{c}{4300} &
 \multicolumn{2}{c}{4370} &  \multicolumn{2}{c}{4440} & \\
\hline
 1.20 && 21.16 & (1)& 20.95 & (26) & 20.72 & (51) & 20.51 & (76) &
20.27 & (101) &  \\
\cline{1-12}
 1.35 && 21.20 & (6)& 21.05 & (31) & 20.81 & (56)&  20.59 & (81) &
20.31 & (106) &  \\
\cline{1-12}
 1.50  && 21.23 & (11)  & 21.09 & (36)  &  20.85 & (61) & 20.62 &
(86) & 20.34 & (111) && [Fe/H] = $-0.70$ \\
\cline{1-12}
 1.65 && 21.26 & (16) & 21.11 & (41) & 20.87 & (66) & 20.64 & (91)  &
20.36 & (116) & \\
\cline{1-12}
 1.80 && 21.28 & (21) & 21.06 & (46) & 20.98 & (71) & 20.60 & (96)  &
20.38 & (121) & \\
\cline{1-14}
 & & & & & &  & & & & &\\
\cline{1-13}
 1.20 && 21.16 & (2) &  20.96 & (27) & 20.73 & (52)  & 20.51 & (77) &
20.28 & (102) &  \\
\cline{1-12}
 1.35 && 21.20 & (7) & 21.03 & (32)  & 20.80 & (57) & 20.57 & (82) &
20.32 & (107) &  \\
\cline{1-12}
 1.50  && 21.23 & (12)  & 21.06 & (37)  &  20.82 & (62) & 20.60 &
(87) & 20.34 & (112) && [Fe/H] = $-0.50$ \\
\cline{1-12}
 1.65 && 21.26 & (17)  & 21.08 & (42) & 20.84 & (67) & 20.62 & (92) &
20.37 & (117) & \\
\cline{1-12}
 1.80 && 21.28 & (22) & 21.06 & (47) & 20.83 & (72) & 20.61 & (97) &
20.54 & (122) & \\
\cline{1-14}
 & & & & & &  & & & & &\\
\cline{1-14}
 1.20 && 21.16 & (3) &  20.96 & (28) & 20.73 & (53)  & 20.52 & (78) &
20.28 & (103) &  \\
\cline{1-12}
 1.35 && 21.20 & (8) & 21.01 & (33)  & 20.78 & (58) & 20.56 & (83) &
20.32 & (108) &  \\
\cline{1-12}
 1.50  && 21.23 & (13)  & 21.04 & (38)  &  20.81 & (63) & 20.59 &
(88) & 20.35 & (113) && [Fe/H] = $-0.30$  \\
\cline{1-12}
 1.65 && 21.26 & (18)  & 21.06 & (43) & 20.83 & (68) & 20.61 & (93) &
20.37 & (118) & \\
\cline{1-12}
 1.80 && 21.27 & (23) & 21.06 & (48) & 20.83 & (73) & 20.78 & (98) &
20.40 & (123) & \\
\cline{1-14}
 & & & & & & & & & & & \\
\cline{1-14}
 1.20 && 21.16 & (4) &  20.96 & (29) & 20.74 & (54)  & 20.52 & (79) &
20.29 & (104) &  \\
\cline{1-12}
 1.35 && 21.20 & (9) & 21.00 & (34)  & 20.77 & (59) & 20.55 & (84) &
20.32 & (109) &  \\
\cline{1-12}
 1.50  && 21.23 & (14)  & 21.02 & (39)  &  20.79 & (64) & 20.57 &
(89) & 20.35 & (114) && [Fe/H] = $-0.15$  \\
\cline{1-12}
 1.65 && 21.25 & (19)  & 21.04 & (44) & 20.82 & (69) & 20.60 & (94) &
20.38 & (119) & \\
\cline{1-12}
 1.80 && 21.27 & (24) & 21.06 & (49)  & 20.84 & (74) & 20.62 & (99) &
20.40 & (124) & \\
\cline{1-14}
 & & & & & &  & & & & &\\
\cline{1-14}
 1.20 && 21.16 & (5) & 20.97 & (30) & 20.74 & (55)  & 20.52 & (80) &
20.28 & (105) &   \\
\cline{1-12}
 1.35 && 21.20 & (10) & 21.00 & (35) & 20.77 & (60)  & 20.55 & (85) &
20.33 & (110) &   \\
\cline{1-12}
 1.50  &&  21.23 & (15) & 21.02 & (40)  & 20.79 & (65)  & 20.58 &
(90) & 20.36 & (115)&&[Fe/H] = $0.00$  \\
\cline{1-12}
 1.65 &&  21.2 & (20) &  21.04 & (45) &  20.82 & (70) & 20.60 & (95)
& 20.38 & (120) &  \\
\cline{1-12}
 1.80 &&  21.27 & (25) &  21.07 & (50) & 20.84 & (75) & 20.63 & (100)
& 20.41 & (125) &  \\
\hline
\hline
\end{tabular}
\end{center}
\end{table*}
\addtolength{\tabcolsep}{-1mm}

\citet{Decin2000A&A...364..137D}  derived an initial value for
$\Omega=$(\Teff\,=\,4320 $\pm$ 140\,K, $\log$ g\,=\,1.50 $\pm$
0.15\,dex, and [Fe/H]\,=\,$-0.50 \pm$ 0.20\,dex), where the
uncertainties on the derived parameters were guessed from (a)
intrinsic uncertainties on the spectra (i.e., the ability to
distinguish between different synthetic spectra at a specific
resolution), (b) the quality of the data, (c) the values of the
non-local Kolmogorov-Smirnov test statistic, and (d) the discrepancies
between observational and synthetic spectra. As such, the estimated
model parameters and their uncertainties in \citet{Decin2000A&A...364..137D}
for the ISO-SWS data are {\em model-dependent external} values.  We merely
use these results to define the values for our grid parameters and for
their spacing.

Let us now properly include both statistical and systematic
 observational errors using the Bayesian approach.  This will enable
 definition of a parameter range for \Teff, $\log$ g, and [Fe/H], and
 selection of the optimal model within the model ensemble specified.
 The analysis will take points (a)--(d) into account in a
 mathematically principled way, providing us with model-dependent
 (error) estimates. How to calculate {\em internal model-dependent error
 estimates} is the subject of Sect.~\ref{internal}. In addition, the
 uncertainty about the model itself, reflected in the so-called
 {\em between-model variability}, is accounted for and combined with the
 internal, or model-dependent, variability, thus producing a measure
 of total variability. Simultaneously accounting for both sources
 properly reflects the true variability and hence produces standard
 errors wider than those obtained, for example, by
 \citet{Griffin1999AJ....117.2998G}. This is extremely important to
 avoid the risk of basing conclusions on noise rather than on signal.

For each model, an MCMC simulation (see Sect.~\ref{posterior_example})
with 10,000 iterations, the first 5000 of which used as burn-in, was
used to calculate the posterior mean of $\mu$ and
$T^{(m)}(y,\theta)$.  { Indeed, when applying MCMC, one typically accounts for the fact that the sequence takes some time before converging to the true posterior distribution by discarding its initial portion \citep{Gilks1996}. When in doubt as to how many iterates should be chopped off, it is prudent to choose a relatively high number.} The variance functions are smoothed with linear mixed models and predicted values used for analysis.

To facilitate comparison with the frequentist results of
\citet{Decin2004A&A...421..281D}, the ranks listed in subsequent
tables and figures are in accordance with the rebinned band 1A data of
$\alpha$ Boo, used by these authors.

\subsection{Determination of stellar parameter ranges}\label{parranges}
Results for the best ten models, as well as for the models which ranked 15,
25, 50, 75, 100, and 125, are given in Table \ref{table1a}. 
\addtolength{\tabcolsep}{2mm}
\begin{table}
\caption{\em \label{table1a}Measures for the goodness-of-fit $T_N$ for some
selected models. The model was estimated using the predicted value of the
linear mixed model for the variance functions. The expected loss values $G(m)$ are given in units of $10^6$. The ranks are chosen for ease of reference to \citet{Decin2004A&A...421..281D}.}
\begin{center}
\tabcolsep=4pt
\begin{tabular}[htb]{ccccrrc}\hline\hline 
& & & & & & Expected\\ 
Rank& Model& \Teff & $\log$ g &\multicolumn{1}{c}{[Fe/H]} &\multicolumn{1}{c}{$T_{N}$} &loss $G(m)$ \\
\hline
1 &62 &4300 &1.50 &$-0.50$ &491.0 &3.403\\
2 &38 &4230 &1.50 &$-0.30$ &490.1 &3.394\\
3 &82 &4370 &1.35 &$-0.50$ &493.2 &3.395\\
4 &61 &4300 &1.50 &$-0.70$ &494.3 &3.403\\
5 &58 &4300 &1.35 &$-0.30$ &495.0 &3.395\\
6 &41 &4230 &1.65 &$-0.70$ &495.9 &3.420\\
7 &102 &4440 &1.20 &$-0.50$ &499.6 &3.405\\
8 &14 &4160 &1.50 &$-0.15$ &497.4 &3.413\\
9 &42 &4230 &1.65 &$-0.50$ &502.0 &3.422\\
10 &81 &4370 &1.35 &$-0.70$ &503.8 &3.422\\
\hline
15 &86 &4370 &1.50 &$-0.70$ &508.1 &3.469\\
25 &15 &4160 &1.50 &$ 0.00$ &515.1 &3.487\\
50 &9 &4160 &1.35 &$-0.15$ &566.4 &3.607\\
75 &11 &4160 &1.50 &$-0.70$ &684.8 &3.970\\
100&117 &4440 &1.65 &$-0.50$ &937.4 &4.995\\
125&125 &4440 &1.80 &$0.00$ &1144.0 &5.728\\
\hline
\hline
\end{tabular}
\end{center}
\end{table}
\addtolength{\tabcolsep}{-2mm} Model 38 has lowest $T^{(m)}(y,\mu)$
value (\Teff\,=\,4230\,K, $\log$ g\,=\,1.50\,dex,
[Fe/H]\,=\,$-0.30$\,dex) with $T^{(38)}(y,\mu)=490.1$. Model 125
(\Teff\,=\,4440\,K, $\log$ g\,=\,1.80\,dex, [Fe/H]\,=\,$-0.00$\,dex)
has the highest value with $T^{(125)}(y,\mu)=1144.0$. Posterior means
as calculated using (\ref{reduce4a}) and $95\,\%$ credible intervals, the Bayesian analog to confidence intervals,
are presented in Fig.~\ref{fig6}.  Fig.~\ref{fig5} shows the
density estimate for the posterior distribution of
$T^{(m)}(y,\mu)$. The density of $T^{(81)}(y,\mu)$, ranking 10th with
\Teff\,=\,4370\,K, $\log$ g\,=\,1.35\,dex, [Fe/H]\,=\,$-0.70$\,dex, is
located to the right, relative to the densities of the other top five
models, underscoring a goodness-of-fit superior to that of model 81,
even though the 95\,\% credible intervals do overlap.  The
model-dependent parameter ranges as estimated from the top 10 models
in our Bayesian analysis range between 4160 and 4440\,K for the
effective temperature, between 1.20 and 1.65\,dex for the logarithm of
the gravity and between $-0.70$ and $-0.15$\,dex for the
metallicity. It will be shown in Sect.~\ref{internal} that the
variability reflected in such ranges can usefully be combined with the
internal error to produce relevant measures of total variability,
meaning that the variability which would follow if the true model were
known is combined with variability resulting from uncertainty about
the model itself.  Note that, by using the frequentist approach of
\citet{Decin2004A&A...421..281D}, the same set of models was selected using
the band 1A ISO-SWS data of $\alpha$ Boo, i.e., the inclusion of the
systematic and statistical errors in the (Bayesian) analysis does not
lead to different parameter ranges. This point is taken up in the
Discussion.
\begin{figure}
\begin{center}
\resizebox{0.39\textwidth}{!}{\rotatebox{0}{\includegraphics{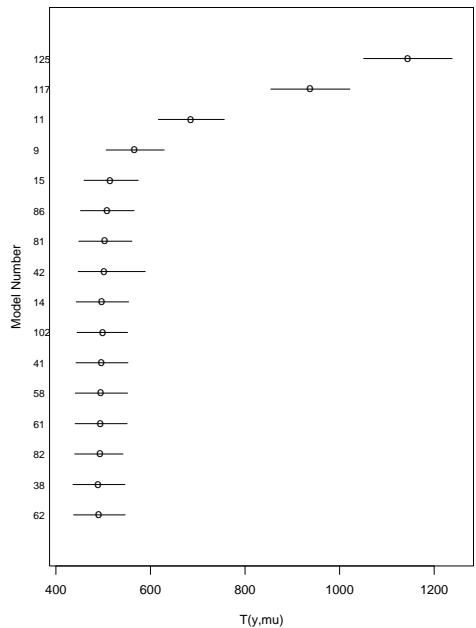}}}
\caption{\em Posterior means and $95\%$ credible intervals for $T(y,\mu)$ for 12 models in band 1A.}
\label{fig6}
\end{center}
\end{figure}
\begin{figure}
\begin{center}
\resizebox{0.42\textwidth}{!}{\rotatebox{0}{\includegraphics{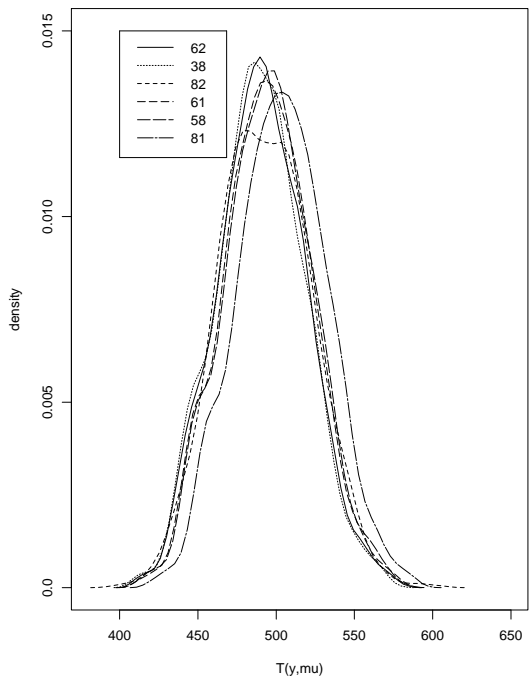}}}
\caption{\em Kernel density estimate for the posterior distribution of
$T^{(m)}(y,\mu)$.}
\label{fig5}
\end{center}
\end{figure}

\subsection{Expected squared error loss} \label{exploss}
\begin{figure}
\begin{center}
\resizebox{0.45\textwidth}{!}{\rotatebox{0}{\includegraphics{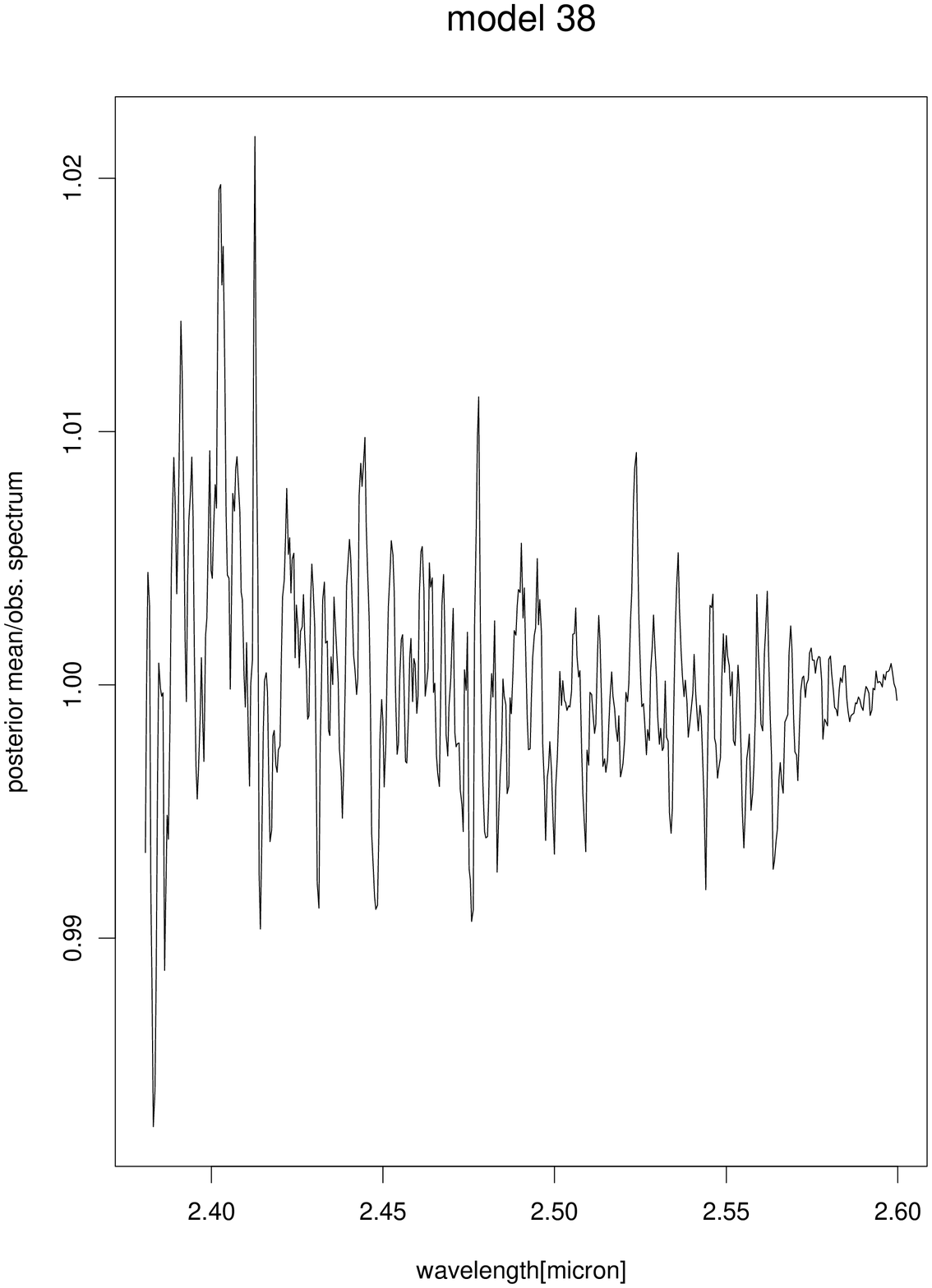}}}
\caption{\em Model 38. Ratio of the posterior mean for the 
synthetic spectrum of model 38 to the observed spectrum of $\alpha$ Boo (\Teff\,=\,4230\,K, $\log$
 g\,=\,1.50\,dex, [Fe/H]\,=\,$-0.30$\,dex) and
posterior mean for the spectrum (full line) in band 1A.}
\label{fig2}
\end{center}
\end{figure}
\begin{figure}
\begin{center}
\resizebox{0.45\textwidth}{!}{\rotatebox{0}{\includegraphics{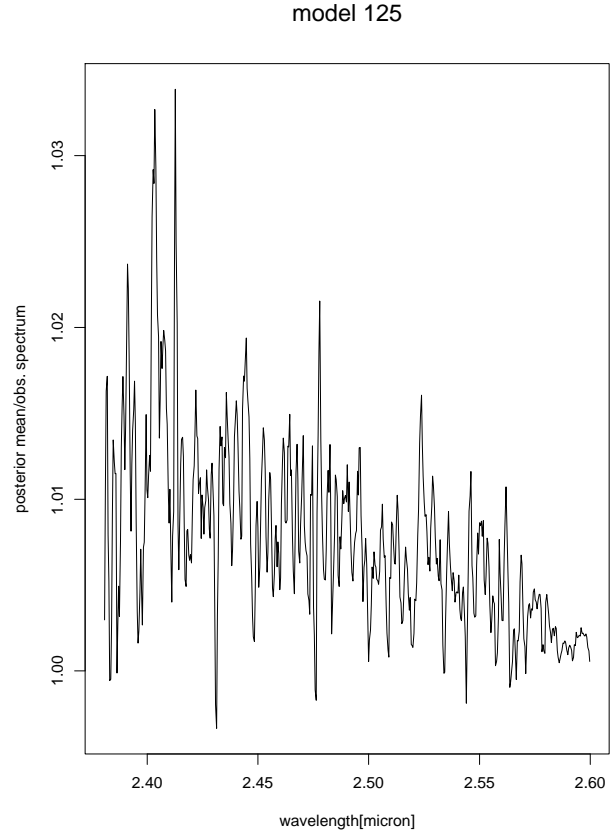}}}
\caption{\em Model 125. Ratio of the posterior mean for the 
synthetic spectrum of model 125 to the observed spectrum of $\alpha$ Boo (\Teff\,=\,4440\,K,
$\log$ g\,=\,1.80\,dex, [Fe/H]\,=\,0.00\,dex) and posterior mean 
for the spectrum (full line) in band 1A.} \label{fig3}
\end{center}
\end{figure}
Model 38 (\Teff\,=\,4230\,K, $\log g\,=\,1.50\,$ dex,
 [Fe/H]\,=\, $-0.30$ \,dex) has the smallest value for
 $L_{38}^{2}$=$3.394 \times 10^{6}$ while model 125 reaches the
 highest value, $L_{125}^{2}=5.7828 \times 10^{6}$. Figs.~\ref{fig2}
 and \ref{fig3} show the observed spectrum, the synthetic spectrum,
 and the posterior mean calculated from (\ref{reduce4a}), for models
 38 and 125.  For model 38, the posterior mean and the observed
 spectrum closely agree along the entire wavelength range. The
 discrepancies are larger for model 125. Note how the posterior mean
 for the spectrum always lies between the observed and synthetic
 spectra. It is also clear for both models that the observed spectrum
 is more dominant at the end of band 1A. Especially for model 125
 (Fig.~\ref{fig3}), the posterior mean and the observed spectrum
 become closer when approaching the end of the band. Based on this
 model selection criterion, model 38 with stellar parameters
 \Teff\,=\,4230\,K, $\log$ g\,=\,1.50\,dex and [Fe/H]\,=\,$-0.30$\,dex
 is selected as providing the best representation of the band 1A
 ISO-SWS data of $\alpha$ Boo.

\subsection{Determination of confidence intervals \label{internal}}
\begin{figure}
\begin{center}
\resizebox{0.4\textwidth}{!}{\rotatebox{0}{\includegraphics{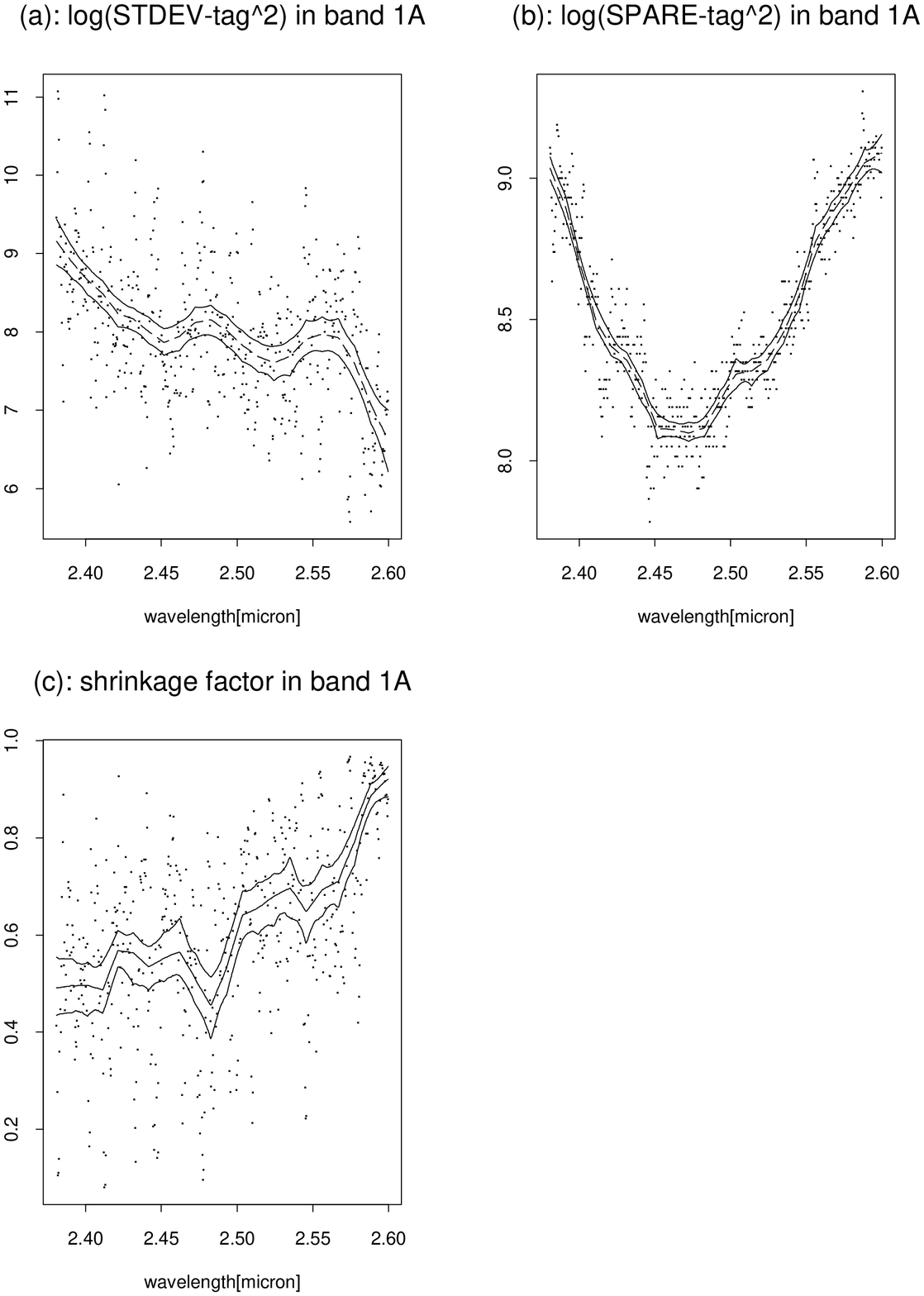}}}
\caption{\em Variance functions. The models were fitted by applying a
linear mixed model for the data.
Panel a: $\log$(STDEV-tag)$^{2}$ in band 1A with the estimated model
and $95$\,\% credible intervals. 
Panel b: $\log$(SPARE-tag)$^{2}$ in band 1A with the estimated model
and 95\,\% credible intervals.
Panel c: shrinkage factor in band 1A with the estimated
model and $95$\,\% credible intervals. }
\label{fig1b}
\end{center}
\end{figure}
Fig.~\ref{fig1b} shows the posterior means and the $95\%$ credible
intervals for $\log(\sigma^{2})$ and $\log(\sigma^{2}_{M})$, as well
as the shrinkage factor determined by a linear mixed model.  The
variance function for both $\sigma$ and $\sigma_M$ is substituted into
the hierarchical model.

As was explained in Sects.~\ref{syntdata} and \ref{posterior}, we
had to restrict calculation of the synthetic spectra to a well-defined
grid, with spacing determined by the analysis of
\citet{Decin2000A&A...364..137D}. However, as we only have the $L^2_m$
values for the predefined grid points, the accuracy of the derived
parameter range for \Teff, $\log$ g, and [Fe/H] is bounded by the grid
spacing. To estimate the confidence intervals around the stellar
parameters and to test the sensitivity of the stellar parameters to
2.38--2.60\,\mic\ IR data of $\alpha$ Boo, we have constrained the
choice of the stellar model and its descriptive parameters by
investigating the behaviour of interpolated stellar models. This kind
of procedure was also followed by \citet{Griffin1999AJ....117.2998G}, who
have simulated a non-linear analytic function to the interpolated
model flux for the purpose of a (frequentist) least-square analysis.

 We chose not to interpolate between the {\em synthetic spectra\/} in
 the grid, but rather to calculate the stratification of a {\em
 theoretical atmosphere model\/} of intermediate mass, gravity or
 effective temperature by interpolating between theoretical models in
 the existing grid, and then to compute the corresponding synthetic
 spectrum. One may argue that for the type of medium-resolution
 spectra we are dealing with the difference between the two
 approaches, i.e., interpolation between the synthetic spectra of the
 existing grid versus computation of new synthetic spectra from
 interpolated theoretical model structures of intermediate $\Omega$,
 will be negligible.  However, our purpose is to develop a general
 tool which, for example, may also be used for observed
 high-resolution spectra. Additionally, since spectral lines behave very
 non-linearly due to saturation, blending, complex dependency on the
 (molecular) opacities for cool-star atmospheres,\dots\ interpolating
 between synthetic spectra should be avoided. For the purpose of the
 interpolation between the models, the quantities as $T$
 (temperature), $\log$ P$_e$ (electron pressure), $\log$ P$_g$ (gas
 pressure), $\log$ a$_{\rm{rad}}$ (radiative acceleration), and $\log
 \kappa$ (extinction coefficient) were interpolated linearly on $\log$
 g or [Fe/H] \citep[see e.g.\ ][]{Plez1992A&AS...94..527P}. To
 interpolate in \Teff, the temperature distribution
 ${\rm{T_{new}}}(\tau)$ was scaled as ${\rm{T_{new}}(\tau) =
 (T_{eff}^{new}/ T_{eff}^{old}) * T_{old}(\tau)}$, followed by a
 pressure integration to calculate the proper P$_e$, P$_g$,\dots  To
 judge upon the accuracy, we have interpolated between
 \Teff\,=\,4230\,K and 4370\,K to obtain \Teff\,=\,4300\,K, between
 $\log$ g\,=\,1.35 and 1.65 to obtain $\log$ g\,=\,1.50, and between
 [Fe/H]\,=\,$-0.30$ and $-0.70$ to obtain [Fe/H]\,=\,$-0.50$ and have
 compared the interpolated model structures (and resulting synthetic
 spectra) with the existing models (and spectra) from the grid. The
 largest difference occurs for the model with the interpolated
 metallicity ([Fe/H]\,=\,$-0.50$) augmenting to 5\,\% for P$_g$ at the
 outermost layer of the atmosphere model. This however only yields a
 discrepancy between the original theoretical spectrum and the one
 calculated from this interpolated model of maximum 0.1\,\% for a
 resolution of $~1500$ (while for a high-resolution spectrum of
 $\Delta \lambda = 0.5$\,\AA, this augments to 0.55\,\%), proving the
 accuracy of our interpolation. Subsequently, we performed a
 1-dimensional interpolation for the parameter values $\Omega$ of the
 selected top 10 models. The parameter spacing for the interpolated
 grid was $\Delta {\rm{T_{eff}}}$\,=\,5\,K, $\Delta log$
 g\,=\,0.01\,dex, and $\Delta$[Fe/H]\,=\,0.01\,dex. Synthetic spectra
 for these interpolated $\Omega$ were then computed.

{ Two comments are in place. First, the reduction of the number of models to the best 10 is not an intrinsic feature of the Bayesian method. Rather, having conducted the aforementioned frequentist analyses, such knowledge can be incorporated into the Bayesian analysis by way of expert priors. In addition, the choice for interpolation is not intrinsically linked to the Bayesian method neither, but rather should be viewed as one of the building blocks of our proposed method.}

Confidence intervals for each of the three parameters were obtained by
calculating the \textit{profile} posterior likelihood for each of the
interpolated models, by holding the two parameters fixed and using the
interpolated grid over the third parameter.  In total, 27 interpolated
models for \Teff, 29 interpolated models for $\log$ g, and 39
interpolated models for [Fe/H] are constructed around one model.  Let,
for example, $G_{k}$, $k=1,\dots,27$, be the profile log-likelihood in
temperature for the $k$th interpolated model. $G_{k}$ is conditioned
upon the values of $\log$ g and [Fe/H]. The normalized profile
log-likelihood is given by
\[
R_{k}=\frac{G_{k}-\min(G_{k})}{\max(G_{k})-\min(G_{k})}.
\]
The interval estimate for \Teff, $\log$ g or [Fe/H] is the set of all
values of \Teff, $\log$ g or [Fe/H] for which the normalized profile
likelihood exceeds 0.9. Table~\ref{table2a} and
Fig.~\ref{figpostlike} exhibit a typical example for determining the
confidence intervals { (here, for model 62, having rank 1 when considering all evidence combined, both provided here and assembled from the literature).} For all of
the top 10 models, the range in the 90\,\% confidence intervals is
$\sim 50$\,K in temperature, $\sim 0.1$\,dex in $\log$ g and $\sim
0.2$\,dex in [Fe/H]. These values thus specify the precision by which
the stellar parameters can be determined, including all sources of
variability.  As a consequence, the best set of stellar parameters
with the associated model-dependent internal error estimates for
$\alpha$ Boo consists of \Teff\,=\,4230\,$\pm$\,25\,K, $\log$
g\,=\,1.50\,$\pm$\,0.05\,dex, and [Fe/H]\,=\,$-0.30 \pm 0.10$\,dex.

Our estimates assume that the model from which they are calculated is
the correct one. Importantly though, this model itself is subject to
uncertainty, illustrated by the fact that not a single model but, say,
10 models (Table~\ref{table1a}) are reasonable
candidates. Constructing ranges from such a collection of models is
useful in its own right, but the information contained therein should
ideally be translated into an additional variance term, to be added to
the internal standard errors. This can formally be done by considering
the total variability surrounding a parameter estimate $\hat{\beta}$:
$$
\mbox{Var}(\hat{\beta})=E[\mbox{Var}(\hat{\beta})|{\cal
M}]+\mbox{Var}[E(\hat{\beta})|{\cal M}],
$$ where ${\cal M}$ represents `model'. The first term on the right
hand side is the internal variance estimate, and is consistently
estimated by the method outlined above. The second term stands for the
variability across models. When choosing, for example, the best 10
models as a representative set, one merely needs to calculate the
sample variance of the corresponding 10 estimates. For \Teff, one
obtains 6312.2, added to $25^2$, yielding 6937.2 and producing an
improved standard error: \Teff\,=\,4230\,$\pm$\,83\,K. For the other
two quantities, the corresponding improved error estimates are: $\log$
g\,=\,1.50\,$\pm$\,0.15\,dex, and [Fe/H]\,=\,$-0.30 \pm 0.21$\,dex.
These error estimates are larger than those obtained by \citet{Griffin1999AJ....117.2998G}, who ignored the between-model variability.
\begin{table}
\caption{\em \label{table2a} Posterior maximum profile likelihood and interval estimates
for \Teff, $\log$ g and [Fe/H] for model 62.}
\begin{center}
\begin{tabular}{lcc}
\hline 
\hline
Parameter & Maximum& (90\,\% C.I.) \\ 
\hline
\Teff & 4295& (4273; 4323) \\
$\log$ g & 1.47 &(1.415; 1.52) \\
${\rm [Fe/H]}$ & $-0.57$& ($-0.67$; $-0.48$) \\
\hline
\hline
\end{tabular}
\end{center}
\end{table}

\begin{figure}
\begin{center}
\resizebox{0.46\textwidth}{!}{\rotatebox{0}{\includegraphics{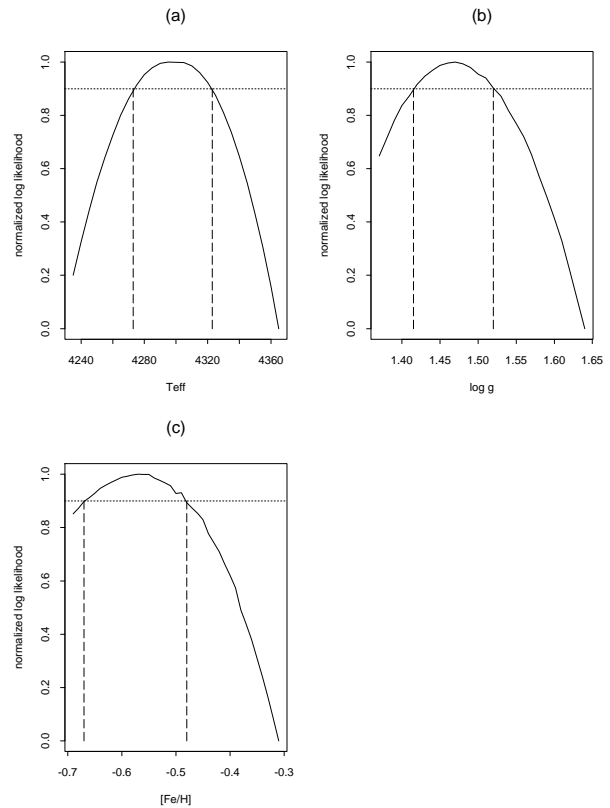}}}
\caption{\em Profile likelihood for model 62. Panel (a): profile
  likelihood for \Teff, panel (b): profile likelihood for $\log$ g,
  and panel (c): profile likelihood for [Fe/H].} 
\label{figpostlike}
\end{center}  
\end{figure}

\section{Discussion} \label{Discussion}
\subsection{Comparison with other statistical methods}
The proposed Bayesian method can
compete with other methods used nowadays for the evaluation of stellar
spectra for deducing stellar parameters. To see this, we discuss in this
section historical work in the same field and analyse sources of involved
errors.

It would indeed be most convincing when we could compare our proposed
Bayesian analysis with other (Bayesian) methods including both
systematic and statistical observational error estimates consistently
throughout the whole analysis of evaluating observational spectra with
theoretical predicitons. However, as far as we are aware of, it
is \emph{the first time} that a statistical method including these
specifications has been developed and used.  Nowadays,
state-of-the-art Bayesian computational techniques are more and more
leaping into the astronomical field, however with the main purpose to
detect a line in a spectral model or a source above background
\citep[][ and references
  therein]{Protassov2002ApJ...571..545P}, to automatically classify
stellar spectra \citep[][]{cheeseman:1996a}, to analyze Poisson
count data \citep{Kraft1991ApJ...374..344K}, to analyze event
arrival times of periodicity \citep{Gregory1992ApJ...398..146G}, to
analyze helioseismology data \citep{Morrow1988IAUS..123..485M}, to
deconvolve astrophysical images \citep{Gull1989}.
\citet{vanDyk1999AAS...194.2604V} were the only 
ones who have employed Bayesian techniques to analyse low-count,
high-resolution astrophysical spectral data. They however have modelled
the source energy spectrum as a mixture of several Gaussian line
profiles and a generalized linear model which accounts for the
continuum, i.e., one assumes that a transformation (e.g., $\log$) of
the model is linear in a set of independent variables, and they have not
computed a full theoretical atmosphere 
model and corresponding synthetic spectrum.

The frequentist approach is the method most often used by astronomers.
The basic approach for modelling data in both the Bayesian and the
frequentist case is the same, the main difference being that a
frequentist route often is more elaborate than its Bayesian
counterpart: (i) one chooses or designs a \emph{figure-of-merit
function} yielding at the end best-fit parameters, (ii) one assesses
the appropriateness of the estimated parameters from a goodness-of-fit
analysis, and (iii) one finally tries to determine the likely errors,
in an {\em ad hoc\/} fashion, for the best-fitting parameters. A few
comments are in place: (1) many practitioners never proceed beyond
item (i), (2) there are numerous instances of inappropriate use of
frequentist methods since practitioners may fail to account for a
method's statistical limitations, calling substantive scientific
results into question \citep[as nicely illustrated by
][]{Protassov2002ApJ...571..545P}, and (3) many statistical methods,
Bayesian and frequentist alike, are designed for use with closed-form
expression. A few examples using this kind of frequentist approach
include \citet{Katz1998A&A...338..151K, Griffin1999AJ....117.2998G,
Cami2000A&A...360..562C, DeBruyne2003MNRAS.339..215D,
Decin2004A&A...421..281D}. A nice example in which a linear regression
method has been developed for the analysis of astronomical data with
measurement errors and intrinsic scatter can be found in
\citet{Akritas1996ApJ...470..706A}. As stated before, two important
conditions made us shift away from frequentist methods: (1) the
inclusion of both statistical and systematic measurement
uncertainties, and (2) the non-availability of a closed analytic
formula to represent the stellar spectrum.

\subsection{On the application to the case-study of $\alpha$ Boo}
{Table~\ref{litaboo} summarizes a comprehensive literature study on the
estimated stellar parameters of our case-study $\alpha$ Boo. A more
elaborate version of this table, listing additionally other parameters such as
the luminosity, the mass, the $^{12}$C/$^{13}$C-ratio, and a short
description of the methods and/or data used by the various authors,
can be found in the appendix of \citet{Decin2000A&A...364..137D}. The table has been updated with the results of \citet{Krticka1999A&AS..138...47K}, \citet{Decin2004A&A...421..281D}, and this study, the only ones using spectrum fitting to determine the stellar parameters for $\alpha$ Boo,
during the past seven years. Provided that error estimates are given
by the authors, they are listed in Table~\ref{litaboo}.
It is clear that many authors do not provide estimates of precision. Second, those who do so typically do not distinguish between the sources of imprecision accounted for, with the noteworthy exception of \citet{Griffin1999AJ....117.2998G} and \citet{Decin2004A&A...421..281D}. These considerations underscore the usefulness of our method. 

\setlength{\tabcolsep}{5mm}

\begin{table*}
\caption{\label{litaboo}Literature study of $\alpha$ Boo: the columns
  tabulate the effective temperature in Kelvin, the
  logaritm of the gravity in cm/s$^2$, and  the
  metallicity, respectively. Values assumed or adopted are given in
  parenthesis. An error estimate is listed whenever provided by the authors.}
\begin{center}
\begin{tabular}{cccl} \hline \hline
\rule[-3mm]{0mm}{8mm}  T$_{\rm{eff}}$ & $\log$ g &  [Fe/H]  & Reference \\ 
\hline
\rule[-0mm]{0mm}{5mm}$4350 \pm 50$ & $1.95 \pm 0.25$ & $-0.5$ & {\citet{vanParadijs1974A&A....35..225V}} \\
$4260 \pm 50$ & $0.90 \pm 0.35$ &  & {\citet{Mackle1975A&AS...19..303M}} \\
$4410 \pm 80$ & & & {\citet{Blackwell1977MNRAS.180..177B}}\\
4240 & &  & {\citet{Linsky1978ApJ...220..619L}} \\
$4060 \pm 150$ & & & {\citet{Scargle1979ApJ...228..838S}} \\
$4420 \pm 150$ & & & {\citet{Blackwell1980A&A....82..249B}} \\
(4260) & (1.6) & & {\citet{Lambert1980ApJ...235..114L}} \\
$4490 \pm 100$ & $2.01 \pm 0.46$ &  $-0.56 \pm 0.07$ & {\citet{Lambert1981ApJ...248..228L}} \\
$4350 \pm 435$ & & & {\citet{Manduca1981ApJ...243..883M}} \\
$4205 \pm 150$ & & & {\citet{Tsuji1981A&A....99...48T}} \\
$4375 \pm 50$ & (1.5) & ($-0.5$) & {\citet{Frisk1982MNRAS.199..471F}} \\
4350 & 1.8 & ($-0.51$) & {\citet{Kjaergaard1982A&A...115..145K}} \\
$4490 \pm 200$ & $2.6 \pm 0.3$ & $-0.55 \pm 0.30$ & {\citet{Burnashev1983IzKry..67...13B}} \\
4370 & & & {\citet{Burnashev1983IzKry..67...13B}} \\
(4375) & (1.57) & &  {\citet{Harris1984ApJ...285..674H}} \\
(4375) & $1.6 \pm 0.2$ &  $-0.5$ &  {\citet{Bell1985MNRAS.212..497B}} \\
($4410$) & ($>0.98$) & ($-0.50$) & {\citet{Gratton1985A&A...148..105G}} \\
($4225$) & $1.6 \pm 0.2$ & ($-0.56$)  & {\citet{Judge1986MNRAS.221..119J}} \\
4400 & 1.7 & $-0.6$ & {\citet{Kyrolainen1986A&AS...65...11K}} \\
(4375) & $1.5 \pm 0.5$ & &  {\citet{Tsuji1986A&A...156....8T}} \\
($4300$) & (1.74) & & {\citet{Altas1987Ap&SS.134...85A}} \\
$4294 \pm 30$ & & & {\citet{DiBenedetto1987A&A...188..114D}} \\
(4375) & $1.97 \pm 0.20$ &  $-0.42$ & {\citet{Edvardsson1988A&A...190..148E}} \\
4321 & (1.8) & ($-0.51$) & {\citet{Bell1989MNRAS.236..653B}} \\
4340 & 1.9 &  $-0.39$ & {\citet{Brown1989ApJS...71..293B}} \\
$4294 \pm 30$ & & & {\citet{Volk1989AJ.....98.1918V}} \\
4300 & 2.0 & $-0.69 \pm 0.10$ & {\citet{Fernandez-Villacanas1990AJ.....99.1961F}} \\
$4280 \pm 200$ & $2.19 \pm 0.27$ &  $-0.60 \pm 0.14$ & {\citet{McWilliam1990ApJS...74.1075M}} \\
$4362 \pm 45$ & & & {\citet{Blackwell1991A&A...245..567B}} \\
$4250 \pm 80$ & $1.6 \pm 0.3$ & & {\citet{Judge1991ApJ...371..357J}} \\
($4375$) & $1.5 \pm 0.5$ & & {\citet{Tsuji1991A&A...245..203T}} \\
4265 & & & {\citet{Engelke1992AJ....104.1248E}} \\
4450 & $1.96 - 1.98$ &  $-0.5$ & {\citet{Bonnell1993MNRAS.264..319B}} \\
4350 & $1.71 - 1.73$ & $-0.5$ & {\citet{Bonnell1993MNRAS.264..319B}} \\
4250 & $1.43 - 1.44$ & $-0.5$ & {\citet{Bonnell1993MNRAS.264..319B}} \\
4250 & $1.81 - 1.82$ & $0.0$ & {\citet{Bonnell1993MNRAS.264..319B}} \\
$4300 \pm 30$ & $1.5 \pm 0.2$ &  $-0.5 \pm 0.1$ & {\citet{Peterson1993ApJ...404..333P}} \\
(4260) & (0.9) & ($-0.77$) & {\citet{Gadun1994AN....315..413G}}\\
(4420) & (1.7)  & ($-0.50$) & {\citet{Gadun1994AN....315..413G}}\\
4362 & 2.4 & & {\citet{Cohen1996AJ....112.2274C}}\\
$4303 \pm 47$ & & & {\citet{Quirrenbach1996A&A...312..160Q}} \\
($4375$) & ($1.5$) &  &  {\citet{Aoki1997A&A...328..175A}} \\
4300 & 1.4 &  $-0.47$ & {\citet{Pilachowski1997AJ....114..819P}}\\
$4291 \pm 48$ & & & {\citet{DiBenedetto1998A&A...339..858D}}\\
4255 & & & {\citet{DiBenedetto1998A&A...339..858D}}\\
$4628 \pm 210$ & & & {\citet{Dyck1998AJ....116..981D}}\\
4320 & & & {\citet{Hammersley1998A&AS..128..207H}} \\
$4321 \pm 44$ & & & {\citet{Perrin1998A&A...331..619P}}\\
 & & $-0.547 \pm 0.021$ & {\citet{Taylor1999A&AS..134..523T}}\\
$4290 \pm 30$ & & &  \citet{Griffin1999AJ....117.2998G}$^{*}$
\\
$4291.9 \pm 0.7$ & $1.94 \pm 0.05$ & $-0.68 \pm 0.02$ &
\citet{Griffin1999AJ....117.2998G}$^{**}$\\ 
$4390 \pm 90$ & $2.0 \pm 0.2$ & $-0.27 \pm 0.05$ &
{\citet{Krticka1999A&AS..138...47K}} \\
$4320 \pm 140$ & $1.50 \pm 0.15$ & $-0.50 \pm 0.20$ & {\citet{Decin2000A&A...364..137D}}\\
4160 -- 4300 & 1.35 -- 1.65 & $-0.30 - 0.00$ & {\citet{Decin2004A&A...421..281D}} \\
\rule[-3mm]{0mm}{3mm}$4230 \pm 83$ &$1.50 \pm 0.15$ & $-0.30 \pm 0.21$ & This paper \\
\hline \hline
\end{tabular}
\end{center}
\footnotesize{$^{*}$: model-independent external errors; $^{**}$:
  model-dependent internal errors} 
\end{table*}

}

Authors using {\em spectroscopic requirements} (i.e., ionisation
  balance, independence of the abundance of an ion versus the
  excitation potential and equivalent width) to estimate the stellar
  parameters for $\alpha$ Boo are
  \citet{vanParadijs1974A&A....35..225V},
  \citet{Mackle1975A&AS...19..303M},
  \citet{Lambert1981ApJ...248..228L}, \citet{Bell1985MNRAS.212..497B},
  \citet{Edvardsson1988A&A...190..148E}, and
  \citet{Bonnell1993MNRAS.264..319B}. {F}rom these results, we infer
  that the values for \Teff\ range between 4260 and 4490\,K, for
  $\log$ g ranging between 0.90 and 2.01\,dex, and for [Fe/H] ranging
  from $-0.56$ to $-0.60$\,dex. The maximum quoted uncertainties are
  100\,K, 0.46\,dex and 0.14\,dex, respectively, although it is not
  always clear whether the authors mention an internal or external
  error estimate.  As has been pointed
  out by, for example, \citet{Smith1985ApJ...294..326S}, one can
  easily assess an external error estimate by varying the derived
  parameter values. This normally results in
  $\Delta$\Teff\,$\approx$\,200K, $\Delta \log$ g\,$\approx$\,0.2\,dex
  and $\Delta$[Fe/H]\,$\approx$\,0.2\,dex.

Only few authors used one or other form of {\em spectrum fitting}
  method to estimate stellar parameters, amongst them
  \citet{Scargle1979ApJ...228..838S}, \citet{Manduca1981ApJ...243..883M},
  \citet{Peterson1993ApJ...404..333P},
  \citet{Krticka1999A&AS..138...47K}, \citet{Decin2000d}, and
  \citet{Decin2004A&A...421..281D}. In the first two of these manuscripts,
  the effective temperature was determined from the flux-curve shape
  alone, while in the others a part of the observational spectrum
  either in the visible or in the near-infrared was used. Values for
  \Teff\ range between 4060 and 4390\,K (with a maximum quoted
  uncertainty of 435\,K from \citet{Manduca1981ApJ...243..883M}), for
  $\log$ g between 1.5 and 2.0\,dex (with maximum uncertainty
  0.2\,dex) and for [Fe/H] between $-0.27$ and $-0.50$\,dex (with
  maximum uncertainty 0.1\,dex).  According to
  \citet{Krticka1999A&AS..138...47K}, the different estimates for the
  stellar parameters as determined from different spectral regions
  using a minimum least-square analysis range between 4200 and 4600\,K
  for \Teff, between 1.53 and 2.35 for $\log$ g and between $-0.155$
  and $-0.461$ for [Fe/H].  \citet{Peterson1993ApJ...404..333P}
  tabulated as results: \Teff\,=\,$4300 \pm 30$\,K, $\log$ g\,=\,$1.5
  \pm 0.2$\,dex, and [Fe/H]\,=\,$-0.5 \pm 0.1$\,dex. We could however
  not trace back if the quoted error estimates include external errors
  or only internal uncertainties.  Only
  \citet{Krticka1999A&AS..138...47K, Decin2000d,
  Decin2004A&A...421..281D} have applied a frequentist least-square
  method to optimize the stellar parameters for $\alpha$ Boo using
  spectrum fitting. None of them included systematic and statistical
  error estimates.

Including both error sources, $\sigma$ and $\sigma_M$ does not result
in different ranges for the fundamental stellar parameters \Teff,
$\log$ g and [Fe/H] of $\alpha$ Boo, relative to
\citet{Decin2004A&A...421..281D}, even though the latter authors did
not take measurement errors into account. Possibly, the error
measurements on the different data points are smaller than the
difference between the observational data and even the best model,
which then would not result in gain of evidence when including the
errors. Comparing the ratio of the observational data to the synthetic
data of model 62 (having rank 1) with $\sigma$ and $\sigma_M$, we note
that all of them have the same order of magnitude. This also indicates
that the remaining structure when considering $y(t)/\theta^{(62)}(t)$,
as in \citet{Decin2004A&A...421..281D}, is not due to measurement
uncertainties but rather indicates that some pattern in the
observational data is not captured by the theoretical
predictions. Plausible explanations for this are: (1) the fact we kept
the C (carbon), N (nitrogen), and O (oxygen) abundance and the
microturbulence fixed, (2) problems with the temperature distribution
in the outermost layers of the model photosphere leading to an
underestimation of the low-level vibration-rotation lines of CO
(carbon monoxide), and (3) problems with the data reduction.

\section{Conclusions and future prospects} \label{conclusions}
Estimating the stellar atmospheric parameters from an observed
 spectrum with given error estimates entails a model selection task in
 which we had to select a synthetic spectrum from a collection of 125
 models. Frequentist methods based on the Kolmogorov-Smirnov test and
 $\chi^{2}$ statistics to assess the goodness-of-fit are unable to
 incorporate the so-called statistical and systematic measurement
 errors of the observational data into the analysis.  Our hierarchical
 Bayesian model with a normal model for the likelihood and conjugate
 normal prior is capable of taking both of these errors into account.
 Using the Bayesian weighted $\chi^2$ statistics to assess the
 goodness-of-fit, the results based on the 2.38--2.60\,$\mu$m ISO-SWS
 data of $\alpha$ Boo are as follows: \Teff\ ranges between 4160 and
 4440\,K, $\log$ g ranges between 1.20 and 1.65\,dex and [Fe/H] ranges
 between $-0.15$ and $-0.70$\,dex. For the model selection process, we
 have used the predictive squared error loss function. The parameters
 of the model with the best representation of the ISO-SWS data are
 \Teff\,=\,4230\,K $\pm 83$\,K), $\log$ g\,=\,1.50\,dex $\pm
 0.15$\,dex), and [Fe/H]\,=\,$-$0.30\,dex $\pm 0.21$\,dex).

Not only here but for a range of applications it is convenient to
first rank the synthetic spectra in the grid, without including
$\sigma$ and $\sigma_M$. When including the observational errors, one
then does not have to apply the Bayesian analysis to all models, like
the 125 considered here, but only to a selection of models that are of
interest, e.g., the models which have the highest ranks and perhaps a
few other models which have a poor goodness-of-fit.
 
It would be of interest, though outside of the scope of this paper, to
apply the proposed method to (1) a larger set of standard stellar
candles analysed in  \citet{Decin2000d}, (2) a 7-dimensional
grid, in which not only the effective temperature, the gravity and the
metallicity are variable, but also the carbon, nitrogen and oxygen
abundance and the microturbulence, and (3) the synthesis analysis of
high-resolution optical data.

We emphasize that the hierarchical Bayesian model as
proposed in this paper is a \emph{general} method which is able to
objectively determine the parameter ranges using the synthesis
technique. In contrast to previous studies, this Bayesian method
incorporates the systematic and statistical measurement error in the
analysis of the data, and so in the determination of the stellar
parameters and their uncertainty intervals. A
  step-by-step algorithmic explanation of the Bayesian analysis
  developed in this paper and the source code thereof are
  available upon request.

\section*{Acknowledgement}
Leen Decin is postdoctoral fellow of the Fund for Scientific Research--Flanders. 
Leen Decin is grateful to Kjell Eriksson for his ongoing
support on the use of the {\sc marcs}-code and to Do Kester for
fruitful discussions on Bayesian analysis.  Ziv Shkedy and Geert
Molenberghs gratefully acknowledge support from the Belgian IUAP/PAI
network ``Statistical Techniques and Modeling for Complex Substantive
Questions with Complex Data". Conny Aerts is supported by the Research
Council of KULeuven under grant GOA/2003/04. 

\mnrasreferences

\label{lastpage}

\begin{thebibliography}{}

\bibitem[\protect\citeauthoryear{{Akritas} \& {Bershady}}{{Akritas} \&
  {Bershady}}{1996}]{Akritas1996ApJ...470..706A}
{Akritas} M.~G.,  {Bershady} M.~A.,  1996, \apj, 470, 706

\bibitem[\protect\citeauthoryear{{Altas}}{{Altas}}{1987}]{Altas1987Ap&SS.134..%
.85A}
{Altas} L.,  1987, \apss, 134, 85

\bibitem[\protect\citeauthoryear{{Aoki} \& {Tsuji}}{{Aoki} \&
  {Tsuji}}{1997}]{Aoki1997A&A...328..175A}
{Aoki} W.,  {Tsuji} T.,  1997, \aap, 328, 175

\bibitem[\protect\citeauthoryear{{Bell}, {Edvardsson} \& {Gustafsson}}{{Bell}
  et~al.}{1985}]{Bell1985MNRAS.212..497B}
{Bell} R.~A.,  {Edvardsson} B.,    {Gustafsson} B.,  1985, \mnras, 212, 497

\bibitem[\protect\citeauthoryear{{Bell} \& {Gustafsson}}{{Bell} \&
  {Gustafsson}}{1989}]{Bell1989MNRAS.236..653B}
{Bell} R.~A.,  {Gustafsson} B.,  1989, \mnras, 236, 653

\bibitem[\protect\citeauthoryear{{Blackwell}, {Lynas-Gray} \&
  {Petford}}{{Blackwell} et~al.}{1991}]{Blackwell1991A&A...245..567B}
{Blackwell} D.~E.,  {Lynas-Gray} A.~E.,    {Petford} A.~D.,  1991, \aap, 245,
  567

\bibitem[\protect\citeauthoryear{{Blackwell}, {Petford} \&
  {Shallis}}{{Blackwell} et~al.}{1980}]{Blackwell1980A&A....82..249B}
{Blackwell} D.~E.,  {Petford} A.~D.,    {Shallis} M.~J.,  1980, \aap, 82, 249

\bibitem[\protect\citeauthoryear{{Blackwell} \& {Shallis}}{{Blackwell} \&
  {Shallis}}{1977}]{Blackwell1977MNRAS.180..177B}
{Blackwell} D.~E.,  {Shallis} M.~J.,  1977, \mnras, 180, 177

\bibitem[\protect\citeauthoryear{{Bonnell} \& {Bell}}{{Bonnell} \&
  {Bell}}{1993}]{Bonnell1993MNRAS.264..319B}
{Bonnell} J.~T.,  {Bell} R.~A.,  1993, \mnras, 264, 319

\bibitem[\protect\citeauthoryear{{Brown}, {Sneden}, {Lambert} \&
  {Dutchover}}{{Brown} et~al.}{1989}]{Brown1989ApJS...71..293B}
{Brown} J.~A.,  {Sneden} C.,  {Lambert} D.~L.,    {Dutchover} E.~J.,  1989,
  \apjs, 71, 293

\bibitem[\protect\citeauthoryear{{Burnashev}}{{Burnashev}}{1983}]{Burnashev198%
3IzKry..67...13B}
{Burnashev} V.~I.,  1983, Izvestiya Ordena Trudovogo Krasnogo Znameni Krymskoj
  Astrofizicheskoj Observatorii, 67, 13

\bibitem[\protect\citeauthoryear{{Cami}, {Yamamura}, {de Jong}, {Tielens},
  {Justtanont} \& {Waters}}{{Cami} et~al.}{2000}]{Cami2000A&A...360..562C}
{Cami} J.,  {Yamamura} I.,  {de Jong} T.,  {Tielens} A.~G.~G.~M.,  {Justtanont}
  K.,    {Waters} L.~B.~F.~M.,  2000, \aap, 360, 562

\bibitem[\protect\citeauthoryear{{Carlin} \& {Louis}}{{Carlin} \&
  {Louis}}{1996}]{Carlin1996}
{Carlin} B.~P.,  {Louis} T.~A.,  1996, Bayes and Empirical methods for data
  analysis.
Chapman and Hall/ CRC, London

\bibitem[\protect\citeauthoryear{Cheeseman \& Stutz}{Cheeseman \&
  Stutz}{1996}]{cheeseman:1996a}
Cheeseman P.,  Stutz J.,  1996, in , Advances in Knowledge Discovery and Data
  Mining.
{AAAI/MIT Press}, pp 153--180

\bibitem[\protect\citeauthoryear{{Cohen}, {Witteborn}, {Carbon}, {Davies},
  {Wooden} \& {Bregman}}{{Cohen} et~al.}{1996}]{Cohen1996AJ....112.2274C}
{Cohen} M.,  {Witteborn} F.~C.,  {Carbon} D.~F.,  {Davies} J.~K.,  {Wooden}
  D.~H.,    {Bregman} J.~D.,  1996, \aj, 112, 2274

\bibitem[\protect\citeauthoryear{{de Bruyne}, {Vauterin}, {de Rijcke} \&
  {Dejonghe}}{{de Bruyne} et~al.}{2003}]{DeBruyne2003MNRAS.339..215D}
{de Bruyne} V.,  {Vauterin} P.,  {de Rijcke} S.,    {Dejonghe} H.,  2003,
  \mnras, 339, 215

\bibitem[\protect\citeauthoryear{{de Graauw}, {Haser}, {Beintema}, {Roelfsema},
  {van Agthoven} \& {et al.}}{{de Graauw}
  et~al.}{1996}]{deGraauw1996A&A...315L..49D}
{de Graauw} T.,  {Haser} L.~N.,  {Beintema} D.~A.,  {Roelfsema} P.~R.,  {van
  Agthoven} H.,    {et al.} 1996, \aap, 315, L49

\bibitem[\protect\citeauthoryear{{Decin}, {Shkedy}, {Molenberghs}, {Aerts} \&
  {Aerts}}{{Decin} et~al.}{2004}]{Decin2004A&A...421..281D}
{Decin} L.,  {Shkedy} Z.,  {Molenberghs} G.,  {Aerts} M.,    {Aerts} C.,  2004,
  \aap, 421, 281

\bibitem[\protect\citeauthoryear{{Decin}, {Vandenbussche}, {Waelkens}, {Decin},
  {Eriksson}, {Gustafsson}, {Plez}, {Sauval} \& {Van Assche}}{{Decin}
  et~al.}{2003}]{Decin2000d}
{Decin} L.,  {Vandenbussche} B.,  {Waelkens} C.,  {Decin} G.,  {Eriksson} K.,
  {Gustafsson} B.,  {Plez} B.,  {Sauval} A.~J.,    {Van Assche} W.,  2003,
  \aap, 400, 709


\bibitem[\protect\citeauthoryear{{Decin}, {Waelkens}, {Eriksson}, {Gustafsson},
  {Plez}, {Sauval}, {Van Assche} \& {Vandenbussche}}{{Decin}
  et~al.}{2000}]{Decin2000A&A...364..137D}
{Decin} L.,  {Waelkens} C.,  {Eriksson} K.,  {Gustafsson} B.,  {Plez} B.,
  {Sauval} A.~J.,  {Van Assche} W.,    {Vandenbussche} B.~D.,  2000, \aap, 364,
  137

\bibitem[\protect\citeauthoryear{{di Benedetto}}{{di
  Benedetto}}{1998}]{DiBenedetto1998A&A...339..858D}
{di Benedetto} G.~P.,  1998, \aap, 339, 858

\bibitem[\protect\citeauthoryear{{di Benedetto} \& {Rabbia}}{{di Benedetto} \&
  {Rabbia}}{1987}]{DiBenedetto1987A&A...188..114D}
{di Benedetto} G.~P.,  {Rabbia} Y.,  1987, \aap, 188, 114

\bibitem[\protect\citeauthoryear{{Dyck}, {van Belle} \& {Thompson}}{{Dyck}
  et~al.}{1998}]{Dyck1998AJ....116..981D}
{Dyck} H.~M.,  {van Belle} G.~T.,    {Thompson} R.~R.,  1998, \aj, 116, 981

\bibitem[\protect\citeauthoryear{{Edvardsson}}{{Edvardsson}}{1988}]{Edvardsson%
1988A&A...190..148E}
{Edvardsson} B.,  1988, \aap, 190, 148

\bibitem[\protect\citeauthoryear{{Engelke}}{{Engelke}}{1992}]{Engelke1992AJ...%
.104.1248E}
{Engelke} C.~W.,  1992, \aj, 104, 1248

\bibitem[\protect\citeauthoryear{{Fern\'{a}ndez-Villaca\~{n}as}, {Rego} \&
  {Cornide}}{{Fern\'{a}ndez-Villaca\~{n}as}
  et~al.}{1990}]{Fernandez-Villacanas1990AJ.....99.1961F}
{Fern\'{a}ndez-Villaca\~{n}as} J.~L.,  {Rego} M.,    {Cornide} M.,  1990, \aj,
  99, 1961

\bibitem[\protect\citeauthoryear{{Frisk}, {Nordh}, {Olofsson}, {Bell} \&
  {Gustafsson}}{{Frisk} et~al.}{1982}]{Frisk1982MNRAS.199..471F}
{Frisk} U.,  {Nordh} H.~L.,  {Olofsson} S.~G.,  {Bell} R.~A.,    {Gustafsson}
  B.,  1982, \mnras, 199, 471

\bibitem[\protect\citeauthoryear{{Gadun}}{{Gadun}}{1994}]{Gadun1994AN....315..%
413G}
{Gadun} A.~S.,  1994, Astronomische Nachrichten, 315, 413

\bibitem[\protect\citeauthoryear{{Gelfand} \& {Ghosh}}{{Gelfand} \&
  {Ghosh}}{1998}]{Gelfand1998}
{Gelfand} A.~E.,  {Ghosh} A.~K.,  1998, Biometrika, 85, 1

\bibitem[\protect\citeauthoryear{{Gelman}, {Carkin}, {Stern} \&
  {Rubin}}{{Gelman} et~al.}{1995}]{Gelman1995}
{Gelman} A.,  {Carkin} J.~B.,  {Stern} H.~S.,    {Rubin} D.~B.,  1995, Bayesian
  data analysis.
Chapman and Hall, London

\bibitem[\protect\citeauthoryear{{Gilks}, {Richardson} \&
  {Spiegelhalter}}{{Gilks} et~al.}{1996}]{Gilks1996}
{Gilks} W.~R.,  {Richardson} S.,    {Spiegelhalter} D.~J.,  1996, Markov Chain
  Monte Carlo in Practice.
Chapman and Hall, London

\bibitem[\protect\citeauthoryear{{Gratton}}{{Gratton}}{1985}]{Gratton1985A&A..%
.148..105G}
{Gratton} R.~G.,  1985, \aap, 148, 105

\bibitem[\protect\citeauthoryear{{Gregory} \& {Loredo}}{{Gregory} \&
  {Loredo}}{1992}]{Gregory1992ApJ...398..146G}
{Gregory} P.~C.,  {Loredo} T.~J.,  1992, \apj, 398, 146

\bibitem[\protect\citeauthoryear{{Griffin} \& {Lynas-Gray}}{{Griffin} \&
  {Lynas-Gray}}{1999}]{Griffin1999AJ....117.2998G}
{Griffin} R.~E.~M.,  {Lynas-Gray} A.~E.,  1999, \aj, 117, 2998

\bibitem[\protect\citeauthoryear{Gull}{Gull}{1989}]{Gull1989}
 Gull, S.F.  1989, in Skilling J.,  ed., {Maximum Entropy and Bayesian Methods}
  {Developments in Maximum Entropy Data Analysis}.
p.~53


\bibitem[\protect\citeauthoryear{{Hammersley}, {Jourdain de Muizon}, {Kessler},
  {Bouchet}, {Joseph}, {Habing}, {Salama} \& {Metcalfe}}{{Hammersley}
  et~al.}{1998}]{Hammersley1998A&AS..128..207H}
{Hammersley} P.~L.,  {Jourdain de Muizon} M.,  {Kessler} M.~F.,  {Bouchet} P.,
  {Joseph} R.~D.,  {Habing} H.~J.,  {Salama} A.,    {Metcalfe} L.,  1998,
  \aaps, 128, 207

\bibitem[\protect\citeauthoryear{{Harris} \& {Lambert}}{{Harris} \&
  {Lambert}}{1984}]{Harris1984ApJ...285..674H}
{Harris} M.~J.,  {Lambert} D.~L.,  1984, \apj, 285, 674

\bibitem[\protect\citeauthoryear{{Judge}}{{Judge}}{1986}]{Judge1986MNRAS.221..%
119J}
{Judge} P.~G.,  1986, \mnras, 221, 119

\bibitem[\protect\citeauthoryear{{Judge} \& {Stencel}}{{Judge} \&
  {Stencel}}{1991}]{Judge1991ApJ...371..357J}
{Judge} P.~G.,  {Stencel} R.~E.,  1991, \apj, 371, 357

\bibitem[\protect\citeauthoryear{{Katz}, {Soubiran}, {Cayrel}, {Adda} \&
  {Cautain}}{{Katz} et~al.}{1998}]{Katz1998A&A...338..151K}
{Katz} D.,  {Soubiran} C.,  {Cayrel} R.,  {Adda} M.,    {Cautain} R.,  1998,
  \aap, 338, 151

\bibitem[\protect\citeauthoryear{{Kessler}, {Steinz}, {Anderegg}, {Clavel},
  {Drechsel}, {Estaria}, {Faelker}, {Riedinger}, {Robson}, {Taylor} \& {Ximenez
  de Ferran}}{{Kessler} et~al.}{1996}]{Kessler1996A&A...315L..27K}
{Kessler} M.~F.,  {Steinz} J.~A.,  {Anderegg} M.~E.,  {Clavel} J.,  {Drechsel}
  G.,  {Estaria} P.,  {Faelker} J.,  {Riedinger} J.~R.,  {Robson} A.,  {Taylor}
  B.~G.,    {Ximenez de Ferran} S.,  1996, \aap, 315, L27

\bibitem[\protect\citeauthoryear{{Kj{\ae}rgaard}, {Gustafsson}, {Walker} \&
  {Hultqvist}}{{Kj{\ae}rgaard} et~al.}{1982}]{Kjaergaard1982A&A...115..145K}
{Kj{\ae}rgaard} P.,  {Gustafsson} B.,  {Walker} G. A.~H.,    {Hultqvist} L.,
  1982, \aap, 115, 145

\bibitem[\protect\citeauthoryear{{Kraft}, {Burrows} \& {Nousek}}{{Kraft}
  et~al.}{1991}]{Kraft1991ApJ...374..344K}
{Kraft} R.~P.,  {Burrows} D.~N.,    {Nousek} J.~A.,  1991, \apj, 374, 344

\bibitem[\protect\citeauthoryear{{Krti{\v c}ka} \& {{\v S}tefl}}{{Krti{\v c}ka}
  \& {{\v S}tefl}}{1999}]{Krticka1999A&AS..138...47K}
{Krti{\v c}ka} J.,  {{\v S}tefl} V.,  1999, \aaps, 138, 47

\bibitem[\protect\citeauthoryear{{Kyr\"{o}l\"{a}inen}, {Tuominen}, {Vilhu} \&
  {Virtanen}}{{Kyr\"{o}l\"{a}inen}
  et~al.}{1986}]{Kyrolainen1986A&AS...65...11K}
{Kyr\"{o}l\"{a}inen} J.,  {Tuominen} I.,  {Vilhu} O.,    {Virtanen} H.,  1986,
  \aaps, 65, 11

\bibitem[\protect\citeauthoryear{{Lambert}, {Dominy} \& {Sivertsen}}{{Lambert}
  et~al.}{1980}]{Lambert1980ApJ...235..114L}
{Lambert} D.~L.,  {Dominy} J.~F.,    {Sivertsen} S.,  1980, \apj, 235, 114

\bibitem[\protect\citeauthoryear{{Lambert} \& {Ries}}{{Lambert} \&
  {Ries}}{1981}]{Lambert1981ApJ...248..228L}
{Lambert} D.~L.,  {Ries} L.~M.,  1981, \apj, 248, 228

\bibitem[\protect\citeauthoryear{{Laud} \& {Ibrahim}}{{Laud} \&
  {Ibrahim}}{1995}]{Laud1995}
{Laud} P.~W.,  {Ibrahim} J.~G.,  1995, Journal of the Royal Statistical Society
  -- Series B, 57, 247

\bibitem[\protect\citeauthoryear{{Linsky} \& {Ayres}}{{Linsky} \&
  {Ayres}}{1978}]{Linsky1978ApJ...220..619L}
{Linsky} J.~L.,  {Ayres} T.~R.,  1978, \apj, 220, 619

\bibitem[\protect\citeauthoryear{{M{\"a}ckle}, {Griffin}, {Griffin} \&
  {Holweger}}{{M{\"a}ckle} et~al.}{1975}]{Mackle1975A&AS...19..303M}
{M{\"a}ckle} R.,  {Griffin} R.,  {Griffin} R.,    {Holweger} H.,  1975, \aaps,
  19, 303

\bibitem[\protect\citeauthoryear{{Manduca}, {Bell} \& {Gustafsson}}{{Manduca}
  et~al.}{1981}]{Manduca1981ApJ...243..883M}
{Manduca} A.,  {Bell} R.~A.,    {Gustafsson} B.,  1981, \apj, 243, 883

\bibitem[\protect\citeauthoryear{{McWilliam}}{{McWilliam}}{1990}]{McWilliam199%
0ApJS...74.1075M}
{McWilliam} A.,  1990, \apjs, 74, 1075

\bibitem[\protect\citeauthoryear{{Morrow} \& {Brown}}{{Morrow} \&
  {Brown}}{1988}]{Morrow1988IAUS..123..485M}
{Morrow} C.~A.,  {Brown} T.~M.,  1988, in IAU Symp. 123: Advances in Helio- and
  Asteroseismology {A Bayesian Approach to Ridge Fitting in the Omega-K Diagram
  of the Solar 5-MINUTE Oscillations}.
pp 485--489.

\bibitem[\protect\citeauthoryear{{Perrin}, {Coude Du Foresto}, {Ridgway},
  {Mariotti}, {Traub}, {Carleton} \& {Lacasse}}{{Perrin}
  et~al.}{1998}]{Perrin1998A&A...331..619P}
{Perrin} G.,  {Coude Du Foresto} V.,  {Ridgway} S.~T.,  {Mariotti} J.~.,
  {Traub} W.~A.,  {Carleton} N.~P.,    {Lacasse} M.~G.,  1998, \aap, 331, 619

\bibitem[\protect\citeauthoryear{{Peterson}, {Dalle Ore} \&
  {Kurucz}}{{Peterson} et~al.}{1993}]{Peterson1993ApJ...404..333P}
{Peterson} R.~C.,  {Dalle Ore} C.~M.,    {Kurucz} R.~L.,  1993, \apj, 404, 333

\bibitem[\protect\citeauthoryear{{Pilachowski}, {Sneden}, {Hinkle} \&
  {Joyce}}{{Pilachowski} et~al.}{1997}]{Pilachowski1997AJ....114..819P}
{Pilachowski} C.,  {Sneden} C.,  {Hinkle} K.,    {Joyce} R.,  1997, \aj, 114,
  819

\bibitem[\protect\citeauthoryear{{Plez}}{{Plez}}{1992}]{Plez1992A&AS...94..527%
P}
{Plez} B.,  1992, \aaps, 94, 527

\bibitem[\protect\citeauthoryear{{Protassov}, {van Dyk}, {Connors}, {Kashyap}
  \& {Siemiginowska}}{{Protassov} et~al.}{2002}]{Protassov2002ApJ...571..545P}
{Protassov} R.,  {van Dyk} D.~A.,  {Connors} A.,  {Kashyap} V.~L.,
  {Siemiginowska} A.,  2002, \apj, 571, 545

\bibitem[\protect\citeauthoryear{{Quirrenbach}, {Mozurkewich}, {Buscher},
  {Hummel} \& {Armstrong}}{{Quirrenbach}
  et~al.}{1996}]{Quirrenbach1996A&A...312..160Q}
{Quirrenbach} A.,  {Mozurkewich} D.,  {Buscher} D.~F.,  {Hummel} C.~A.,
  {Armstrong} J.~T.,  1996, \aap, 312, 160

\bibitem[\protect\citeauthoryear{{Scargle} \& {Strecker}}{{Scargle} \&
  {Strecker}}{1979}]{Scargle1979ApJ...228..838S}
{Scargle} J.~D.,  {Strecker} D.~W.,  1979, \apj, 228, 838


\bibitem[\protect\citeauthoryear{{Shkedy}}{{Shkedy}}{2003}]{Zivthesis}
{Shkedy} Z.,  2003, PhD thesis, Limburgs Universitair Centrum, Belgium

\bibitem[\protect\citeauthoryear{{Smith} \& {Lambert}}{{Smith} \&
  {Lambert}}{1985}]{Smith1985ApJ...294..326S}
{Smith} V.~V.,  {Lambert} D.~L.,  1985, \apj, 294, 326

\bibitem[\protect\citeauthoryear{{Taylor}}{{Taylor}}{1999}]{Taylor1999A&AS..13%
4..523T}
{Taylor} B.~J.,  1999, \aaps, 134, 523

\bibitem[\protect\citeauthoryear{{Tsuji}}{{Tsuji}}{1981}]{Tsuji1981A&A....99..%
.48T}
{Tsuji} T.,  1981, \aap, 99, 48

\bibitem[\protect\citeauthoryear{{Tsuji}}{{Tsuji}}{1986}]{Tsuji1986A&A...156..%
..8T}
{Tsuji} T.,  1986, \aap, 156, 8

\bibitem[\protect\citeauthoryear{{Tsuji}}{{Tsuji}}{1991}]{Tsuji1991A&A...245..%
203T}
{Tsuji} T.,  1991, \aap, 245, 203

\bibitem[\protect\citeauthoryear{{van Dyk}, {Kashyap}, {Siemiginowska} \&
  {Conners}}{{van Dyk} et~al.}{1999}]{vanDyk1999AAS...194.2604V}
{van Dyk} D.~A.,  {Kashyap} V.~L.,  {Siemiginowska} A.,    {Conners} A.,  1999,
  Bulletin of the American Astronomical Society, 31, 864

\bibitem[\protect\citeauthoryear{{van Paradijs} \& {Meurs}}{{van Paradijs} \&
  {Meurs}}{1974}]{vanParadijs1974A&A....35..225V}
{van Paradijs} J.,  {Meurs} E. J.~A.,  1974, \aap, 35, 225

\bibitem[\protect\citeauthoryear{{Verbeke} \& {Molenberghs}}{{Verbeke} \&
  {Molenberghs}}{2000}]{Verbeke2000}
{Verbeke} G.,  {Molenberghs} G.,  2000, Linear mixed models for longitudinal
  data.
Springer New-York

\bibitem[\protect\citeauthoryear{{Volk} \& {Cohen}}{{Volk} \&
  {Cohen}}{1989}]{Volk1989AJ.....98.1918V}
{Volk} K.,  {Cohen} M.,  1989, \aj, 98, 1918

\end{thebibliography}
\end{document}